\documentclass[journal]{IEEEtran}
\newcommand{\beq}{\begin{equation}}
\newcommand{\eeq}{\end{equation}}
\newcommand{\bsq}{\begin{subequations}}
\newcommand{\esq}{\end{subequations}}
\newcommand{\bq}{\begin{eqnarray}}
\newcommand{\eq}{\end{eqnarray}}
\newcommand{\bqn}{\begin{eqnarray*}}
\newcommand{\eqn}{\end{eqnarray*}}

\DeclareMathAlphabet{\mathcal}{OMS}{cmsy}{m}{n}         
\usepackage{helvet}         
\usepackage{courier}        
\usepackage{type1cm}        
\usepackage{makeidx}         
\usepackage{graphicx}        

\usepackage{multicol}        
\usepackage[bottom]{footmisc}
\usepackage{ifthen}
\usepackage{subfigure}
\usepackage[usenames,dvipsnames]{color}

\makeindex             

\usepackage{graphics} 
\usepackage{graphicx} 
\usepackage{epsfig} 
\usepackage{epstopdf}
\usepackage{mathptmx} 
\usepackage{times} 
\usepackage[cmex10]{amsmath}

\usepackage{mathtools}  
\usepackage{amssymb}  
\usepackage{amsthm}

\usepackage{amsfonts}
\usepackage{cite}
\usepackage{verbatim}
\usepackage{comment}
\usepackage{algorithmic}
\usepackage{multirow}
\usepackage{cite}
\usepackage{stfloats}
\usepackage{supertabular}
\usepackage{longtable}

\DeclareGraphicsExtensions{.pdf,.png,.jpg,.eps,.ps}
\renewcommand{\arraystretch}{1.2}
\usepackage{arydshln}
\usepackage{multicol}
\usepackage{colortbl}
\theoremstyle{definition}
\newtheorem{theorem}{Theorem}
\newtheorem{lemma}{Lemma}

\newtheorem{proposition}{Proposition}

\theoremstyle{definition}
\newtheorem{definition}{Definition}

\ifCLASSINFOpdf
\else
\fi

\usepackage{comment}
\usepackage{ifthen}
\newboolean{showcomments}
\setboolean{showcomments}{true}
\newcommand{\ychen}[1]{\ifthenelse{\boolean{showcomments}}
        { \textcolor{red}{(YC:  #1)}}{}}
\newcommand{\fliu}[1]{\ifthenelse{\boolean{showcomments}}
        { \textcolor{blue}{(FL:  #1)}}{}}
\newcommand{\slow}[1]{\ifthenelse{\boolean{showcomments}}
        { \textcolor{green}{(SL:  #1)}}{}}

\usepackage{amsmath}
\allowdisplaybreaks[4]

\usepackage[ruled]{algorithm2e}

\begin{document}

%
\title{An Energy Sharing Mechanism Considering Network Constraints and Market Power Limitation}
%
%
%

\author{Yue Chen,
		Changhong Zhao, Steven H. Low,
		and Adam Wierman
	\thanks{Y. Chen is with the Department of Mechanical and Automation Engineering, the Chinese University of Hong Kong, HKSAR, China. (e-mail: yuechen@mae.cuhk.edu.hk)}
	\thanks{C. Zhao is with the Department of Information Engineering, the Chinese University of Hong Kong, HKSAR, China. (email: chzhao@ie.cuhk.edu.hk)}
	\thanks{S. Low and A. Wierman are with the Computing + Mathematical Sciences Department, California Institute of Technology, Pasadena, CA 91125 USA (e-mails: \{slow, adamw\}@caltech.edu)}
\thanks{This work has been submitted to the IEEE for possible publication.Copyright may be transferred without notice, after which this version may no longer be accessible.}}
	

%
%

\markboth{Journal of \LaTeX\ Class Files,~Vol.~XX, No.~X, Feb.~2019}%
{Shell \MakeLowercase{\textit{et al.}}: Bare Demo of IEEEtran.cls for IEEE Journals}
%



\maketitle

\begin{abstract}
As the number of prosumers with distributed energy resources (DERs) grows, the conventional centralized operation scheme may suffer from conflicting interests, privacy concerns, and incentive inadequacy. In this paper, we propose an energy sharing mechanism to address the above challenges. It takes into account network constraints and fairness among prosumers. In the proposed energy sharing market, all prosumers play a generalized Nash game. The market equilibrium is proved to have nice features in a large market or when it is a variational equilibrium. To deal with the possible market failure, inefficiency, or instability in general cases, we introduce a price regulation policy to avoid market power exploitation. The improved energy sharing mechanism with price regulation can guarantee existence and uniqueness of a socially near-optimal market equilibrium. Some advantageous properties are proved, such as prosumer's individual rationality, a sharing price structure similar to the locational marginal price, and the tendency towards social optimum with an increasing number of prosumers. For implementation, a practical bidding algorithm is developed with convergence condition. Experimental results validate the theoretical outcomes and show the practicability of our model and method.
\end{abstract}

\begin{IEEEkeywords}
prosumers, networked energy sharing, generalized Nash equilibrium, price regulation, bidding algorithm
\end{IEEEkeywords}

%
\IEEEpeerreviewmaketitle

\section*{Nomenclature}
\addcontentsline{toc}{section}{Nomenclature}
\subsection{Abbreviations}	\begin{IEEEdescription}[\IEEEusemathlabelsep\IEEEsetlabelwidth{${\underline P _{m}}$,${\overline P _{m}}$}]
	\item[DER] Distributed energy resource.
	\item[GNE] Generalized Nash equilibrium.
	\item[GNG] Generalized Nash game.
	\item[LMP] Locational marginal price.
	\item[P2P] Peer-to-peer.
	\item[PV] Photovoltaic.
	\item[PoA] Price of anarchy.
\end{IEEEdescription}
\subsection{Indices, Sets, and Functions}
	\begin{IEEEdescription}[\IEEEusemathlabelsep\IEEEsetlabelwidth{${\underline P _{m}}$,${\overline P _{m}}$}]
		\item[$i, \mathcal{I}$] Index and set of prosumers.
		\item[$l,\mathcal{L}$] Index and set of lines.
		\item[$S_i$] Action set of prosumer $i$, and  $S=\prod_{i \in \mathcal{I}}S_i$.
		\item[$J_i(p_i)$] Cost or disutility function of prosumer $i$, and $J(p)=\sum_{i \in \mathcal{I}} J_i(p_i)$.
		\item[$\Gamma_i(p,b)$] Total cost of prosumer $i$ with energy sharing, which equals $J_i(p_i)+\lambda_i(-a\lambda_i+b_i)$. 
		\item[$u_i(b,p_i)$] Payment to the market under the improved energy sharing mechanism.
	\end{IEEEdescription}
	\subsection{Parameters}
	\begin{IEEEdescription}[\IEEEusemathlabelsep\IEEEsetlabelwidth{${\underline P _{m}}$,${\overline P _{m}}$}]
		\item[$I$] Number of prosumers.
		\item[$p_i^0$] Original energy production of prosumer $i$.
		\item[$E_i^0$] Energy purchased from the main grid or aggregator of prosumer $i$.
		\item[$D_i^0$] Fixed demand of prosumer $i$.
		\item[$D_i$] Energy reduction of prosumer $i$.
		\item[$c_i,d_i$] Coefficients of function $J_i(p_i)$ for prosumer $i$.
		\item[$a$] Energy sharing market sensitivity.
		\item[$\hat F_l$] Power flow limit for line $l$; $\tilde F_l:=\hat F_l+\sum_{i=1}^I \pi_{il}(E_i^0-D_i)$ and $F_l:=\hat F_l-\sum_{i=1}^I \pi_{il}(E_i^0-D_i)$.
		\item[$\pi_{il}$] Line flow distribution factor from bus $i$ to line $l$.
	\end{IEEEdescription}
		\subsection{Decision Variables}
		\begin{IEEEdescription}[\IEEEusemathlabelsep\IEEEsetlabelwidth{${\underline P _{m}}$,${\overline P _{m}}$}]
		\item[$p_i$] Production adjustment of prosumer $i$.
		\item[$\lambda_i$] Energy sharing price for prosumer $i$.
		\item[$\lambda_i^r$] Regulated energy sharing price for prosumer $i$.
		\item[$q_i$] Amount of energy prosumer $i$ gets from sharing.
		\item[$b_i$] Bid of prosumer $i$ in the energy sharing market.
		\item[$\eta, \alpha_l^{\pm}$] Dual variables of problem \eqref{eq:new-rule}.
		\item[$\beta_i$] Dual variables of problem \eqref{eq:sharing-game}.
		\item[$\kappa,\tau_l^{\pm}$] Dual variables of problem \eqref{eq:central}.
		\end{IEEEdescription}



        
\section{Introduction}
\label{sec-1}
%
%
%
%

 \IEEEPARstart{T}{he} advance in distributed generation technologies and the decline in costs of electrical devices encourage the continuous integration of distributed energy resources (DERs) such as rooftop photovoltaic (PV) panels, small wind turbines, and electric vehicles \cite{akorede2010distributed}. The distinct merits of DERs in environmental friendliness and flexibility render them a promising role in constructing a green smart grid \cite{salinas2013dynamic}. However, challenges come along: On the one hand, the proliferation of DERs induces the transformation of traditional passive consumers to proactive ``prosumers''\cite{parag2016electricity}. Prosumers are equipped with renewable generators and have more flexible means for energy management by altering their production and consumption. This results in two-sided uncertainties from fluctuating renewable generations and unpredictable demands. On the other hand, DERs are typically operated by different stakeholders who make individual decisions and possess private information. The conflicting interests, information asymmetry, and two-sided uncertainty lead to supply-demand mismatch that can jeopardize system security, raise DER operating costs, and stymie the DER integration process \cite{ryu2020real}.


Over decades, the centralized scheme, where users purchase electricity from the aggregator/retailer, and the aggregator/retailer sells (buys) surplus (insufficient) electricity to (from) the grid, has been proved to be effective \cite{wei2014energy}. However, in a prosumer era, the conventional centralized scheme lacks efficiency in managing proactive participants and facilities due to the aforementioned challenges. Moreover, the low feed-in tariffs hinder the integration of renewable energy. Therefore, an innovative business model that can provide adequate incentives is desired \cite{sousa2019peer}. Inspired by recent prosperity of sharing economy \cite{Benjaafar2018Peer} in transportation, lodging, etc., \emph{energy sharing} \cite{chen2018analyzing} becomes a promising concept given its potential in smoothing uncertainty \cite{ryu2020real}, reducing peak demand \cite{tushar2019grid}, and so on.

The initial form of energy sharing runs with the help of a central operator, which coordinates all the distributed devices, makes best matches, and allocates the profits directly or indirectly via price setting. 
The energy sharing among PV prosumers  was modeled as a Stackelberg game \cite{liu2017energy,Sharing-setprice2}, in which the microgrid operator acts as a leader and the prosumers act as followers. Existence and uniqueness of the Stackelberg equilibrium were proved in \cite{cui2017distributed}. Prosumers' flexibility in subscribing to different energy sharing regions was considered in \cite{chen2020peer}. Uncertain electricity prices and volatile renewable outputs were taken into account in \cite{xu2020two}, which used a descent algorithm to search for the equilibrium. Heterogeneous preferences of market participants on source of energy \cite{morstyn2018multiclass} and risk \cite{moret2020heterogeneous} were considered. A data-driven approach based on the spatial-temporal graph convolutional networks was proposed to characterize the preference of prosumers \cite{chen2021data}.
The studies above exploited prices as intermediary; alternatively, profit allocation can also be set in advance through a contract between the operator and participants. 
Cooperative game theory was adopted to incentivize prosumer coalitions \cite{han2018incentivizing,mei2019coalitional}. Reference \cite{chakraborty2018sharing} compared the case in which prosumers are equipped with and willing to share storage versus the case in which prosumers would like to invest in a joint storage. 
Shapley value is a common tool for splitting profits within a coalition, which can be estimated using the stratified random sampling method \cite{han2019estimation} and extended to discrete case \cite{Albizuri2014Monotonicity}. K-means clustering was applied to reduce computational burden and improve scalability of energy sharing \cite{han2019improving}. The initial form of energy sharing mechanisms reviewed above is consistent with the current ``aggregator/retailer-user'' operating structure. However, the main obstacle for implementation lies in fetching prosumers' private information to design a proper and fair pricing or allocation policy. This initial form may also restrict initiatives of prosumers merely as price-takers.

To overcome the limitations above, a broad literature turned to exploring active roles of prosumers as price-makers. Work along this line allows market participants to bid either the quantity or a function of quantity and price. The former type is called a Cournot competition in economics. The potential efficiency loss in a Cournot competition was revealed \cite{tsitsiklis2014efficiency}. The latter type is more similar to the actual electricity market operation where step-wise offering functions are submitted by the generators. A supply function bidding method was introduced in a demand response program \cite{li2015demand}, and reference \cite{xu2015demand} further incorporated capacity constraints. Another parameterized supply function proposed in \cite{johari2011parameterized} was proved to minimize the worst-case welfare loss within a class of market mechanisms. Apart from the pool market, bilateral contracts were studied in \cite{morstyn2018bilateral} by matching the sellers to the buyers. The above work divides the participants into buyers and sellers beforehand. This limits prosumers’ flexibility since their market roles as buyers or sellers are in fact changeable in the market. Reference \cite{chen2020energy} proposed a generalized demand bidding mechanism for node-level energy sharing and proved properties of its market equilibrium; \cite{chen2020approaching} further provided a practical bidding process. Reference \cite{kalathil2017sharing} modeled the storage investment decision of firms with sharing options as a non-cooperative game, for which a unique equilibrium exists under a mild condition. Despite the efforts above, energy trading among price-making participants has not been fully explored partly because of the sophisticated models involved, such as a generalized Nash game or an equilibrium problem with equilibrium constraint, especially when network constraints are considered. How an agent decides on the optimal selling quantity in a transmission-constrained Cournot competition was studied in \cite{barquin2008cournot} by assuming a known sensitivity matrix. A method to characterize the residual demand derivative with consideration of network constraints was developed in \cite{xu2007transmission} by enumerating all possible combination of binding constraints. A Cournot competition based mechanism with a price cap was introduced \cite{yao2007two}. Two models were developed for the cases with binding/non-binding price cap constraints, respectively. As for the supply function bidding method, reference \cite{liu2007impacts} revealed that network constraints could result in multiple equilibria or no pure Nash equilibrium. Reference \cite{le2020peer} discussed the equilibria of distributed peer-to-peer markets, and revealed that when agents have no consensus on the value of the same product, the bilateral trade prices may diverge and there is no guarantee of a unique equilibrium. Reference \cite{guerrero2018decentralized} presented a sensitivity-based methodology for P2P trading in a low-voltage network. Though network constraints were considered, the above work either relies on simplifying assumptions or only reveals potential market failures without proposing solutions.

This paper proposes an energy sharing mechanism considering price-making prosumers, network constraints, and endogenously-given market roles. With a well-designed price regulation policy, the proposed mechanism can avoid market power exploitation and ensure the existence of a unique market equilibrium. The main contributions are three-fold:

1) \textbf{Mechanism Design to Incorporate Network Constraints}. We propose an energy sharing mechanism for prosumers considering network constraints. Distinct from previous work, the market platform aims to minimize price discrimination\footnote{Charging prosumers different prices for energy based on what the seller believes the prosumer would agree to.} rather than maximizing its own profit. The proposed mechanism can protect privacy with no need of prosumers’ disutility functions to clear the market and enables the endogenous determination of market roles. We prove that the energy sharing market equilibrium, which is a generalized Nash equilibrium (GNE), exists and is unique in a large market when prosumers are price-takers or when the GNE is a variational equilibrium. However, as illustrated by counterexamples, there is no general guarantee for existence, uniqueness, or optimality of GNE due to market power exploitation, which calls for further improvement as in our Contribution 2).

2) \textbf{Price Regulation to Limit Market Power}. We propose a price regulation policy to avoid market power exploitation. The regulation policy restricts price privilege of every prosumer (compared to its marginal production adjustment cost) to a level depending on its sharing quantity, the market sensitivity, and the total number of prosumers. In this way, we ensure existence and uniqueness of a socially near-optimal GNE. We prove that a Pareto improvement is achieved among prosumers and that the resulting energy sharing price has a similar structure to the locational marginal price (LMP). The total cost of prosumers under energy sharing approaches the social optimum as the number of prosumers grows.

3) \textbf{Bidding Algorithm to Achieve Equilibrium.} For implementation, we introduce a bidding algorithm to achieve the improved energy sharing market equilibrium in a distributed manner. A guidance to select market parameters is provided to assure convergence of the bidding algorithm. Economic intuition of the convergence condition is explained based on cobweb model.

The rest of this paper is organized as follows. Section \ref{sec-2} proposes an energy sharing mechanism considering network constraints; properties of its market equilibrium are discussed in Section \ref{sec-3}, revealing the possibility of market failure, inefficiency, and instability; to overcome this problem, a price regulation policy is presented and proved to be effective in Section \ref{sec-4};
a bidding process to achieve the improved equilibrium is introduced in Section \ref{sec-5}; some possible extensions are discussed in Section \ref{sec:extension}; numerical case studies are carried out in Section \ref{sec-6}; Section \ref{sec-7} concludes the paper. For ease of reading, we summarize our main results below:

1) An energy sharing mechanism for networked prosumers is proposed in Section \ref{sec-2.2}. We prove that a unique equilibrium exists with socially optimal efficiency in a large market with price-taking prosumers in Proposition \ref{Thm:prop-0} or with socially near-optimal efficiency when the GNE is a variational equilibrium in Proposition \ref{Thm:prop-1}. Two counterexamples are given in Section \ref{sec-3.2} showing that however in general cases, there is no guarantee for existence, uniqueness, or optimality of GNE.

2) We introduce a price regulation policy \eqref{eq:priceregulate} in Section \ref{sec-4.1}, giving rise to an improved energy sharing mechanism. The existence and uniqueness of the GNE under the improved mechanism are proved in Theorem \ref{Thm:prop-2}. Some properties of the GNE are revealed: the improved energy sharing mechanism achieves Pareto improvement over self-sufficiency in Proposition \ref{Thm:prop-3}, the price-of-anarchy (PoA) tends to 1 with an increasing number of prosumers in Proposition \ref{Thm:prop-5}, and the energy sharing price adopts a similar structure to the LMP in Proposition \ref{Thm:prop-4}.

3) A practical bidding process is presented in Algorithm 1 with proof of its convergence in Theorem \ref{Thm:prop-7}.

\section{Networked Energy Sharing Mechanism}
\label{sec-2}
In this section, we propose an energy sharing mechanism considering network constraints under which all prosumers play a generalized Nash game.
\subsection{Problem description}
\label{sec-2.1}
A set of networked prosumers indexed by $i \in \mathcal{I}=\{1,2,...,I\}$ is considered. For simplicity, we assume that each prosumer has a distributed generator (DG). The self-energy-balance condition for each prosumer $i \in \mathcal{I}$ is \eqref{eq:self-balance}, where $p_i^0$ is prosumer $i$'s energy production, $E_i^0$ is its energy purchased from the main grid or aggregator, and $D_i^0$ is its demand.
\begin{align}
\label{eq:self-balance}
    p_i^0 + E_i^0 = D_i^0
\end{align}

All the prosumers take part in the interruptible/curtailable program that belongs to the incentive-based demand response \cite{albadi2007demand}. In this type of demand response program, the participant is asked to adjust its demand to a predefined value and will receive payment or penalty based on its performance.
Suppose prosumer $i \in \mathcal{I}$ is required to reduce its energy purchased from the main grid by $D_i$. To meet this requirement, prosumer $i \in \mathcal{I}$ increases its net production (which may be realized by increasing generation or reducing demand) by $p_i$, which becomes $(p_i^0-D_i^0)+p_i$.
This adjustment results in an extra cost or disutility $J_i(p_i):=c_i p_i^2+d_i p_i$ with constant coefficients $c_i>0$ and $d_i$.

Define \emph{self-sufficiency} as the default response that every prosumer $i \in \mathcal{I}$ can only increase its net production to fulfill its desired energy purchase reduction, without trading energy with other prosumers: $(p_i^0-D_i^0+p_i)+(E_i^0-D_i)=0$, i.e., $p_i=D_i$ for all $i \in \mathcal{I}$. 
This self-sufficiency scheme is how the current system operates and is simple since no coordination among the prosumers is needed. But is there a more efficient way? In fact, \emph{energy sharing} can provide a better solution to this problem. By allowing a prosumer with lower marginal disutility to produce more and sell energy to another prosumer with higher marginal disutility, a win-win situation can be reached among them.

Here, we use “sharing” not to refer to the natural pooling effect of the grid according to the Kirchhoff’s laws, but to the sharing of the prosumers’ adjustable capabilities. The questions we ask are: how to design a sharing market (coordination mechanism) to allow the prosumers jointly reduce their aggregate grid purchase by $\sum_{i}D_i$, as requested by the system operator, in a way that satisfies power balance and line limits? What are the optimality and convergence properties of a market equilibrium? By sharing their adjustable capabilities, an individual prosumer $i$ may not reduce its grid purchase by its scheduled amount $D_i$, some reduce more and some less, so that collectively, they provide the required total reduction $\sum_{i}D_i$. We hope that everyone is better off than without the sharing market, and the system approaches social optimality as the number of prosumers grows. 

The remaining problem is how to design such a proper energy sharing mechanism, which is:
1) \emph{Private}. Prosumer information privacy is preserved. 2) \emph{Motivated}. Each prosumer has the incentive to participate in energy sharing. 3) \emph{Effective}. Balance of supply and demand is reached, and physical network constraints are satisfied. 4) \emph{Flexible}. Each prosumer has the freedom to choose between being a seller or a buyer.

\subsection{Market clearing with network constraints}
\label{sec-2.2}
In this paper, we propose an energy sharing mechanism considering network constraints. Denote the amount of energy prosumer $i\in \mathcal{I}$ purchases from the energy sharing market as $q_i$, so that it can meet the energy reduction requirement as $p_i+q_i=D_i$. A generalized demand function \cite{hobbs2000strategic} is used to represent the relationship between $q_i$, prosumer bid $b_i$, and the energy sharing price $\lambda_i$. A different energy sharing price $\lambda_i$ is assigned to each prosumer $i \in \mathcal{I}$ to reflect its influence on power flow across transmission lines. To be specific:

1) The energy sharing amount of prosumer $i \in \mathcal{I}$ is: 
\begin{align}
\label{eq:demand-function}
q_i= -a \lambda_i+b_i 
\end{align}
where $q_i>0$ if prosumer $i$ is a buyer and $q_i<0$ if it is a seller; $a>0$ is the market sensitivity, which measures impact of prosumer bids on the energy sharing price; $b_i$ is the bid of prosumer $i$, showing its willingness to buy. Here, a linear function is used because it well captures the nature of the decrease in purchase/selling quantity with higher/lower prices and facilitates analysis. A linear function was also used in references \cite{li2015demand,xu2015demand}. Note that the demand function can be represented as $q_D=-a_D\lambda+b_D$ with $a_D>0$. The supply function can be represented as $q_S=a_S\lambda-b_S$ with $a_S>0$, which is equivalent to $-q_S=-a_S\lambda+b_S$. For a prosumer, its sensitivity on buying or selling energy should be similar, i.e., $a_D \approx a_S$. Therefore, the demand and supply functions can be consolidated as a uniform form $q_i=-a\lambda_i+b_i$ with $a>0$.

2) Energy balancing condition:
\begin{align}
\label{eq:clearing}
\sum \nolimits_{i=1}^I q_i = \sum \nolimits_{i=1}^I (-a\lambda_i+b_i)=0
\end{align}
It means the amount of energy sold to the energy sharing market equals the amount of energy bought from the market.

3) Power flow limits for the underlying network:
\begin{align}
\label{eq:flowlimits}
-\tilde F_l \le \sum \nolimits_{i=1}^I \pi_{il} (-a\lambda_i+b_i) \le F_l,~\forall l \in \mathcal{L}
\end{align}
The nodal net power injection at node $i$ after sharing energy is $E_i^0-D_i+q_i$. Let $\hat F_l,\forall l$ be the line capacity, then the network constraints are $-\hat F_l \le \sum_{i=1}^I \pi_{il}(E_i^0-D_i+q_i) \le \hat F_l,\forall l \in \mathcal{L}$, which is equivalent to \eqref{eq:flowlimits} with $\tilde F_l:=\hat F_l+\sum_{i=1}^I \pi_{il}(E_i^0-D_i)$ and $F_l:=\hat F_l-\sum_{i=1}^I \pi_{il}(E_i^0-D_i)$.
Direct current (DC) model is used to calculate the power flow on each line $l \in \mathcal{L}=\{1,...,L\}$. $\pi_{il}$ is the line flow distribution factor from prosumer $i$ to line $l$.

\textbf{Remark on line flow distribution factor calculation.} Consider the network as a directed graph with arbitrarily assigned directions. Let $C \in \{0,1,-1\}^{I\times L}$ be the incidence matrix of the network; $B \in (\mathbb{R}^+)^{L\times L}$ be the diagonal matrix with diagonal terms being positive line weights derived from the standard DC power flow model, i.e., for a line $ij \in \mathcal{L}$, its weight $B_{ij} = |V_i||V_j|/x_{ij}$  where $|V_i|$ and $|V_j|$ are voltage magnitudes at nodes (prosumers) $i$ and $j$, respectively, and $x_{ij}$ is the reactance of inductive line $ij$. 
Let $\tilde C$ be a reduced incidence matrix by removing an arbitrary row of $C$ (without loss of generality, we remove the last row). Then the line flow distribution factor matrix $\Pi$ can be constructed as follows: first calculate $\tilde \Pi = -\left(\tilde C B \tilde C^T\right)^{-1} \tilde C B \in \mathbb{R}^{(I-1)\times L}$; then add an all-zero row of dimension $L$ to the bottom of $\tilde \Pi$ to obtain $\Pi$, whose element in $i$-th row, $l$-th column is $\pi_{il}$.

With all prosumers' bids $b_i,\forall i \in \mathcal{I}$, a central platform clears the market to determine the energy sharing quantities $q_i,\forall i \in \mathcal{I}$ and prices $\lambda_i,\forall i \in \mathcal{I}$, satisfying constraints \eqref{eq:clearing}-\eqref{eq:flowlimits}. Though $\lambda_i,\forall i$ and $q_i,\forall i$ are set by the platform, both of them are influenced by the prosumers' bids $b_i,\forall i$ through problem \eqref{eq:new-rule}. Therefore, compared with the case using a set-price, prosumers in our model are in fact more flexible since they can influence both prices and quantities. We design a market clearing rule for the central platform as the solution of the following optimization:
\bsq
\label{eq:new-rule}
\begin{align}
    \mathop{\min}_{\lambda_i,\forall i\in \mathcal{I}} ~&   \sum \nolimits_{i=1}^I \lambda_i^2 \label{eq:new-rule.1} \\
    \mbox{s.t.} ~ & \sum \nolimits_{i=1}^I (-a\lambda_i+b_i)=0~ :\eta \label{eq:new-rule.2}\\
    ~ & -\tilde F_l \le \sum \nolimits_{i=1}^I \pi_{il}(-a\lambda_i+b_i) \le F_l ~:\alpha_l^{\pm},\forall l  \in \mathcal{L} \label{eq:new-rule.3}
\end{align}
\esq

The proposed rule \eqref{eq:new-rule} has several advantages: 

1) \emph{Unique market outcome}. The optimal solution of \eqref{eq:new-rule}, if exists, must be unique because of the strict convexity of objective \eqref{eq:new-rule.1}. Given the prosumers' bids $b_i,\forall i \in \mathcal{I}$ and the unique energy sharing prices $\lambda_i,\forall i \in \mathcal{I}$, the sharing quantities $q_i,\forall i \in \mathcal{I}$ are also unique due to \eqref{eq:demand-function}.

2) \emph{Ensure fairness of the market.} Minimizing the objective function \eqref{eq:new-rule.1} is equivalent to minimizing the variance of all prosumers’ energy sharing prices. This is because
\begin{align}
    ~ & \frac{1}{I} \sum_{i=1}^I \left(\lambda_i-\frac{1}{I}\sum_{i=1}^I \lambda_i\right)^2 = \frac{1}{I} \sum_{i=1}^I \left(\lambda_i-\frac{1}{aI}\sum_{i=1}^I b_i\right)^2 \nonumber\\
    = ~ & \frac{1}{I} \left[\sum_{i=1}^I \lambda_i^2 -\frac{2}{aI}(\sum_{i=1}^I b_i)(\sum_{i=1}^I \lambda_i) +\frac{1}{a^2I^2} (\sum_{i=1}^I b_i)^2 \right] \nonumber\\
    = ~ & \frac{1}{I} \sum_{i=1}^I \lambda_i^2 + \frac{1}{I}(\frac{1}{a^2I^2}-\frac{2}{a^2I})(\sum_{i=1}^I b_i)^2
\end{align}
The first and third equations are due to constraint \eqref{eq:new-rule.2}. In the platform’s problem \eqref{eq:new-rule}, multiplying the objective function \eqref{eq:new-rule.1} by a constant $\frac{1}{I}$ will not affect the optimal solution; the bids $b_i,\forall i \in \mathcal{I}$ are given and fixed, so the term $\frac{1}{I}(\frac{1}{a^2I^2}-\frac{2}{a^2I})(\sum_{i=1}^I b_i)^2$ is a constant.
This enables us to modify \eqref{eq:new-rule.1} to 
$\frac{1}{I}\sum_{i=1}^I \left(\lambda_i - (\sum_{j=1}^I\lambda_j)/I\right)^2$
without affecting its optimal solution over the feasible set \eqref{eq:new-rule.2}--\eqref{eq:new-rule.3}. Hence, problem \eqref{eq:new-rule} essentially reduces price discrimination by minimizing the variance of $\lambda_i$ over $i \in \mathcal{I}$. This ensures a fair market outcome for all prosumers. 

3) \emph{Uniform price without congestion}. When the network constraints \eqref{eq:new-rule.3} are not binding, the optimal solution of \eqref{eq:new-rule} is  $\lambda_i={\sum_{j=1}^I b_j}/{(aI)}$ for all $ i\in \mathcal{I}$, indicating the same price for all prosumers. This aligns with the conventional electricity market setting where all participants have a uniform electricity price when there is no congestion.

4) \emph{Endogenously-given market roles}. The market roles of prosumers are not predertemined but set by the market clearing problem \eqref{eq:new-rule}. Take the case without congestion as an example, the uniform energy sharing price is $\lambda=\sum_{i=1}^I b_j/(aI)$, therefore, $q_i=-a\lambda+b_i = -\sum_{i=1}^I b_j/I +b_i$. When the prosumer $i \in \mathcal{I}$ is more willing to buy than the average, i.e., $b_i>\sum_{i=1}^I b_j/I$, we have $q_i>0$ and it becomes a buyer; otherwise, it becomes a seller. This is a distinct feature of our model that is different from the conventional electricity markets. In wholesale or retail electricity markets, the roles of participants are determined beforehand. Usually, the generator acts as a seller and the load acts as a buyer. The seller and buyer have different bidding rules. In contrast, when prosumers enter the proposed market, their roles are symmetric, and who becomes a seller and who becomes a buyer depend on the situation of others. An example is given in TABLE \ref{tab:marketrole}.


\subsection{Networked energy sharing game}
\label{sec-2.3}
\begin{figure}[!t]
\centerline{\includegraphics[width=0.65\columnwidth]{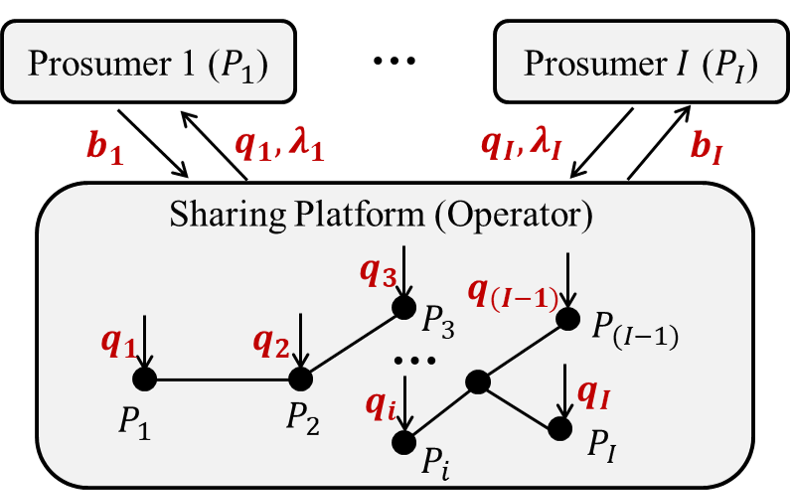}}
\setlength{\abovecaptionskip}{0pt}
\caption{Procedure of the sharing mechanism.}
\label{fig:procedure}
\end{figure}

Fig. \ref{fig:procedure} illustrates the proposed energy sharing mechanism: Each prosumer $i \in \mathcal{I}$ offers a bid $b_i$ to a market platform. Upon receiving all the bids $b:=\{b_i,~\forall i \in \mathcal{I}\}$, the platform solves \eqref{eq:new-rule} to determine price $\lambda_i(b)$ and sharing amount $q_i(b)=-a\lambda_i(b)+b_i$ for all $i \in \mathcal{I}$. If $q_i(b)>0$, prosumer $i$ purchases energy by making payment $\lambda_i(b) q_i(b)=\lambda_i(b) (-a\lambda_i(b)+b_i)$ to the market; otherwise prosumer $i$ sells energy to the market and receives revenue $-\lambda_i(b) q_i(b)$. 

Each prosumer $i \in \mathcal{I}$ aims to minimize its disutility $J_i(p_i) =  c_ip_i^2+d_ip_i$ plus payment (or minus revenue) to (from) the energy sharing market while maintaining energy balance. Formally, for each prosumer $i\in\mathcal{I}$:
\bsq
\label{eq:sharing-game}
\begin{align}
    \mathop{\min}_{p_i,b_i} ~ & c_ip_i^2+d_ip_i + \lambda_i(b)(-a\lambda_i(b)+b_i) \label{eq:sharing-game.1}\\
    \mbox{s.t.}~ & p_i-a\lambda_i(b)+b_i=D_i ~: \beta_i \label{eq:sharing-game.2} 
\end{align}
\esq
where $\lambda_i(b)$ is the $i$-th element of the optimal solution of \eqref{eq:new-rule} parameterized by vector $b$. We denote objective \eqref{eq:sharing-game.1} as $\Gamma_i(p,b)$ and the feasible set defined by the self-energy-balance constraint \eqref{eq:sharing-game.2} as $(p_i, b_i) \in S_i(p_{-i}, b_{-i})$ \footnote{$p_{-i}:=\{p_n,\forall n \in \mathcal{I}, n \ne i \},~ b_{-i}:=\{b_n,\forall n \in \mathcal{I}, n \ne i \}$.}.

In \eqref{eq:sharing-game}, all prosumers constitute a generalized Nash game (GNG) \cite{harker1991generalized} composed of the following elements:

 \noindent 1) The set of players $\mathcal{I}=\{1,2,...,I\}$; 

 \noindent 2) Strategy sets $S_i(p_{-i},\!b_{-i}),\forall i \!\in \!\mathcal{I}$; strategy space $S \!=\! \prod_{i \in \mathcal{I}} \!S_i$; 

\noindent 3) Payoff functions $\Gamma_i(p,b), \forall i \in \mathcal{I}$. 

For simplicity, denote by $\mathcal{G}=\{\mathcal{I},S,\Gamma\}$ the abstract form of GNG \eqref{eq:sharing-game}. The GNG differs from the standard Nash game (SNG) in that not only the objective function but also the strategy set of one player depends on the other players’ strategies \cite{scutari2012monotone}. In this paper, we consider a price-making prosumer, who takes into account the impact of its bid $b_i$ on the market price $\lambda_i$. When clearing the market, coupling constraints such as the network constraints are considered by the platform. The feasible set of $\lambda_i$ depends on the other players’ bids $b_{-i}$. This will further influence the feasible set $S_i$ of prosumer $i$’s action $(p_i,b_i)$ due to the constraint \eqref{eq:sharing-game.2}. Therefore, the proposed energy sharing market is modeled as a GNG instead of a SNG.
Although a general GNG may be intractable, in the next section, we can characterize some useful properties of the equilibrium of the specific GNG \eqref{eq:sharing-game}.

\section{Equilibrium of the Networked Sharing Game}
\label{sec-3}
An equilibrium of the networked energy sharing game above is a generalized Nash equilibrium, formally defined as follows.

\begin{definition} (Generalized Nash Equilibrium)
A strategy profile $ (\bar p,\bar b) \in S $
is a \emph{generalized Nash equilibrium} (GNE) of the networked energy sharing game $\mathcal{G}$ in \eqref{eq:sharing-game}, if for all $i \in \mathcal{I}$:
\begin{align}
    \Gamma_i(\bar p_i, \bar b_i, \bar p_{-i}, \bar b_{-i}) \le \Gamma_i(p_i, b_i, \bar p_{-i}, \bar b_{-i}) , \forall (p_i,b_i) \in S_i(\bar p_{-i}, \bar b_{-i}). \nonumber
\end{align}
\end{definition}

\subsection{Properties of GNE in two special cases}
\label{sec-3.1}
We show that the GNE of the proposed energy sharing game \eqref{eq:sharing-game} has nice properties in two special cases: 1) in a large market with price-taking prosumers; 2) when the GNE happens to be a variational equilibrium. We use the \emph{social optimum} as a benchmark, which is defined as below.

\begin{definition} (Social Optimum)
A point $\tilde p=(\tilde p_1,\dots,\tilde p_I)$ is a \emph{social optimum} or \emph{socially optimal} if it is the unique optimal solution of the centralized operation problem:
\bsq
\label{eq:AGG}
\begin{align}
\mathop{\min}_{p_i,\forall i \in \mathcal{I}}~ & \sum \limits_{i=1}^I \left(c_ip_i^2+d_ip_i\right) \label{eq:AGG.1}\\
\mbox{s.t.} ~ & \sum \limits_{i=1}^I p_i = \sum \limits_{i=1}^I D_i \label{eq:AGG.2}\\
~ & -\tilde F_l \le \sum \limits_{i=1}^I \pi_{il}(D_i-p_i) \le F_l, \forall l\in \label{eq:AGG.3} \mathcal{L}
\end{align}
\esq 
\end{definition}

The social optimum is well defined if and only if the following feasibility condition holds:

\textbf{A1}: $\left\{p \in \mathbb{R}^{I}~|~ p ~\mbox{satisfies}~\eqref{eq:AGG.2}-\eqref{eq:AGG.3}.\right\} \ne \emptyset$.

We first consider the equilibrium in a large market, where there are many prosumers that the impact of each prosumer's strategy on the price vector $\lambda$ can be ignored. In that case, prosumers act as ``price-takers'' where prosumer $i$ solves problem \eqref{eq:sharing-game} with a constant $\lambda_i$ rather than as a function of $b$. 
This motivates the following definition.
\begin{definition} (Competitive Equilibrium) 
A tuple $(\bar p, \bar b, \bar \lambda)$ is a competitive equilibrium (CE) if for all $i \in \mathcal{I}$, given $\bar \lambda_i$, 
\bsq \label{eq:competitiveequilibrium}
\begin{align}
    (\bar p_i, \bar b_i)=\mbox{argmin}_{p_i,b_i} ~ & c_ip_i^2+d_ip_i+\bar\lambda_i (-a\bar \lambda_i+b_i) \\
    \mbox{s.t.}~ & p_i-a\bar\lambda_i+b_i=D_i
\end{align}
\esq
and given the bid $\bar b$, the price $\bar \lambda$ is the optimal solution of \eqref{eq:new-rule}.
\end{definition}

\begin{proposition}
\label{Thm:prop-0}
Suppose A1 holds.  Then a unique CE $(\bar p,\bar b, \bar \lambda)$ of the networked energy sharing game exists.  Moreover $\bar p$ is socially optimal.
\end{proposition}
The proof of Proposition \ref{Thm:prop-0} is in  Appendix \ref{apen-0}. It reveals that the proposed sharing market can achieve the same efficiency as centralized operation (social optimum) in a large market with price-taking prosumers.
Later in Proposition \ref{Thm:prop-5}, we prove that the GNE tends to the CE when the prosumer number $I \to \infty$.
In contrast, when prosumer number $I$ is small, prosumers are price-makers and can exercise market power to manipulate price 
away from the social optimum. 
In this situation, we consider a special case where the GNE happens to be a \emph{variational equilibrium} (VE) \footnote{Variational equilibrium refers to a GNE where the Lagrangian multipliers associated with shared constraints are identical across all the prosumers $i\in\mathcal{I}$. 
}.
The following proposition points out properties of VE without requiring a large market (a large number $I$). 

\begin{proposition}
\label{Thm:prop-1}
Suppose the power network is radial. If a VE $(\bar p,\bar b)$ of the networked energy sharing game $\mathcal{G}$ in \eqref{eq:sharing-game} exists, then it must be the unique VE. Moreover, $\bar p$ is the unique optimal solution of the following problem:
\bsq
\label{eq:central}
\begin{align}
\mathop {\min} \limits_{p_i,\forall i \in \mathcal{I}} ~& \sum \limits_{i=1}^I c_i p_i^2 + d_i p_i +\sum \limits_{i=1}^I \frac{(D_i- p_i)^2}{2a(I-1)} \label{eq:central.1}\\
\mbox{s.t.} ~& \sum \limits_{i=1}^I p_i =\sum \limits_{i=1}^I D_i ~: \kappa \label{eq:central.2}\\
~ & -\tilde F_l \le \sum \limits_{i=1}^I \pi_{il}(D_i- p_i) \le F_l~:\tau_l^{\pm},\forall l\in\mathcal{L} \label{eq:central.3}
\end{align}
\esq
and $\bar b_i= 2ac_i\bar p_i+ad_i+2(D_i-\bar p_i)$ for all $i \in \mathcal{I}$.
\end{proposition}

The proof of Proposition \ref{Thm:prop-1} is in Appendix \ref{apen-1}. We notice that problem \eqref{eq:central} is similar to the social optimum problem \eqref{eq:AGG} except for an extra term in the objective function. This shows that the VE has a socially near-optimal efficiency. Moreover, when the prosumer number $I \to \infty$, the extra term $\sum_{i=1}^I \frac{(D_i-p_i)^2}{2a(I-1)}$ tends to zero and problem \eqref{eq:central} becomes the centralized operation problem \eqref{eq:AGG}, which is consistent with the result for a large market (Proposition \ref{Thm:prop-0}). However, despite the  nice properties above, the existence, uniqueness, and optimality of GNE in Propositions \ref{Thm:prop-0} and \ref{Thm:prop-1} may or may not hold in general. This is demonstrated by two examples below, and is what motivates our improved design of networked energy sharing.

\subsection{Examples illustrating market power exploitation}
\label{sec-3.2}
In the following, we use two examples to show the possible market power exploitation resulting in market failure (no equilibrium), market inefficiency, or market instability (multiple equilibria). Example 1 shows that market instability happens when the transmission line is congested, while Example 2 further introduces a case where the above three adverse consequences of market power exploitation might happen.

\begin{figure}[h]
\centerline{\includegraphics[width=0.5\columnwidth]{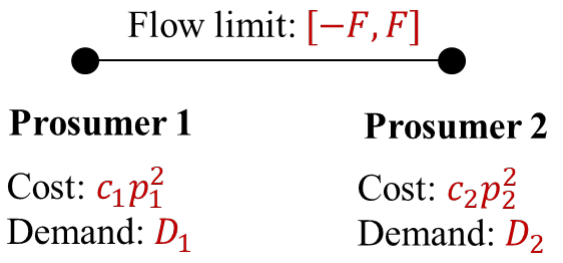}}
\setlength{\abovecaptionskip}{0pt}
\caption{Settings of Example 1 with two prosumers.}
\label{fig:example1}
\end{figure}

\textbf{Example 1}: Two prosumers connect to the head bus and the tail bus of a line, respectively. Set $a=1$, $c_1=c_2=c$, $d_1=d_2=0$, $E_1^0=D_1$, $E_2^0=D_2$, and the line flow limit to be $F$ as in Fig. \ref{fig:example1}.





\emph{Mathematical Analysis:} In this example, network constraint \eqref{eq:new-rule.3} is simply $-F \le  -\lambda_1+b_1 \le F$. Given Prosumer 2's strategy $b_2$, the price $\lambda_1$ solved from \eqref{eq:new-rule} is a function of $b_1$:
\bq
\nonumber
\lambda _1(b_1) = \begin{cases} 
F+b_1 &~ {\mbox{if}} ~ \frac{b_1- b_2}{2}\le -F \\
(b_1+ b_2)/2 &~ {\rm{if}} ~ -F < \frac{b_1- b_2}{2} < F \\
-F+b_1 &~ {\mbox{if}} ~ \frac{b_1- b_2}{2}\ge F 
\end{cases}
\eq
where the first and last scenarios correspond to cases in which the line flow reaches its lower and upper bound, respectively, and in the second scenario the line flow constraint is inactive. Prosumer 1's decision-making problem is thus:
\bq
\label{eq:example2}
\mathop{\min}_{b_1\in \mathbb{R}}~  c(D_1+\lambda_1(b_1)-b_1)^2+\lambda_1(b_1)(-\lambda_1(b_1)+b_1)
\eq

Given $b_2$, it can be verified that Prosumer 1's objective \eqref{eq:example2} is continuous on $b_1 \in\mathbb{R}$; moreover, it is linear and strictly decreasing on $(-\infty,~b_2-2F]$, quadratic and strictly convex on $(b_2 - 2F, ~b_2+2F)$, and linear and strictly increasing on $[b_2+2F, ~+\infty)$. Therefore, given $b_2$, the best response $b_1$ of prosumer 1 depends on the relationship between the axis of symmetry and the boundary points $(b_2\pm 2F)$ of the quadratic segment in its objective function. Specifically:
\begin{eqnarray}
b_1 = 
\begin{cases} 
b_2 + 2F, ~ \textnormal{if}~ b_2 \leq 2cD_1-2(c+1)F  \quad (P1U)\\ 
b_2-2F, ~ \textnormal{if}~ b_2\geq2cD_1+2(c+1)F  \quad (P1L) \\
\frac{c}{c+1}(b_2 +2D_1), ~ \textnormal{otherwise} \quad (P1M)
\end{cases} \nonumber
\end{eqnarray}
Similarly, given $b_1$, the best response of prosumer 2 is:
\begin{eqnarray}
b_2 = 
\begin{cases} 
b_1 + 2F, ~ \textnormal{if}~ b_1 \leq 2cD_2-2(c+1)F  \quad (P2U)\\ 
b_1-2F, ~ \textnormal{if}~ b_1\geq2cD_2+2(c+1)F  \quad (P2L) \\
\frac{c}{c+1}(b_1 +2 D_2), ~ \textnormal{otherwise} \quad (P2M)
\end{cases} \nonumber
\end{eqnarray}
For any GNE $(\bar p, \bar b)$, its $(\bar b_1, \bar b_2)$ must fall in one of the following scenarios \emph{exclusively}: $(P1M)~\&~(P2M)$, or $(P1U)~\&~(P2L)$, or $(P1L)~\&~(P2U)$, while all the other scenarios lead to contradiction. These three scenarios correspond to the three cases of GNEs as shown below. In particular, for scenario $(P1U)~\&~(P2L)$, one must have:
\begin{eqnarray}
2cD_2+2cF \leq \bar b_1 - 2F =  \bar b_2 \leq 2cD_1-2(c+1)F  \nonumber
\end{eqnarray}
Similar conditions can be derived for scenarios $(P1L)~\&~(P2U)$, and $(P1M)~\&~(P2M)$. Therefore,

$\bullet$  If $|D_1-D_2|<(2c+1)F/c$:
\begin{align}
    \bar b_1 =~ & c(D_1+D_2)+\frac{c}{2c+1}(D_1-D_2) \nonumber\\
    \bar b_2 = ~& c(D_1+D_2)+\frac{c}{2c+1}(D_2-D_1) \nonumber\\
    \bar \lambda_1 =~ &  \bar \lambda_2 = c(D_1+D_2) \nonumber
\end{align}
and $\bar p_i = D_i +\bar \lambda_i - \bar b_i$ for $i=1,2$. It can be verified that $(\bar p,\bar b)$ is the unique GNE in this case. Moreover, $(\bar p_1, \bar p_2)$ coincides with the optimal solution of \eqref{eq:central}.

$\bullet$  If $D_1-D_2 \ge (2c+1)F/c$, then every $(\bar p,\bar b)$ that satisfies:
\begin{align}
    \bar b_2 \in ~& \left[2cD_2+2cF,~2cD_1-2(c+1)F\right] \nonumber\\
    \bar b_1  =~ &  \bar b_2 + 2F \nonumber\\
    \bar \lambda_1 =~& -F+\bar b_1; \quad \bar \lambda_2 =  F+ \bar b_2 \nonumber
\end{align}
and $\bar p_i = D_i +\bar \lambda_i - \bar b_i$ for $i=1,2$ is a GNE. Line congestion occurs since $D_1 - \bar p_1 = F = -(D_2-\bar p_2)$, and there are a range of GNEs which have the same $\overline p$ that is optimal for \eqref{eq:central}.

$\bullet$  If $D_1-D_2 
\le -(2c+1)F/c$, then every $(\bar p,\bar b)$ that satisfies:
\begin{align}
    \bar b_2 \in ~& \left[2cD_1+2(c+1)F, ~2cD_2 -2cF\right] \nonumber\\
    \bar b_1  =~ &  \bar b_2 - 2F \nonumber\\
    \bar \lambda_1 =~& F+\bar b_1; \quad \bar \lambda_2 =   -F+ \bar b_2 \nonumber
\end{align}
and $\bar p_i = D_i +\bar \lambda_i - \bar b_i$ for $i=1,2$ is a GNE.  Line congestion occurs since $-(D_1 - \bar p_1) = F = D_2-\bar p_2$, and there are a range of GNEs which have the same $\overline p$ that is optimal for \eqref{eq:central}. $\hfill \qedsymbol$

\emph{Economics Intuition}: In this example, the only difference between two prosumers lies in the required energy reductions $D_1$ and $D_2$. This difference offers room for energy sharing. When $|D_1-D_2|$ is large, no matter which bid they start with, at some point during the bidding process the line will be congested with the power flow fixed. Take Prosumer 1 as an example and suppose $-\lambda_1+b_1$ is fixed to $-F$, then according to \eqref{eq:example2}, it can constantly lower its total cost by offering a larger $b_1$ as long as the line is still congested. When Prosumer 1 increases its bid $b_1$ to an extreme, there is one GNE. With different starting points, there will be multiple GNEs. When the gap $|D_1-D_2|$ is small, the line will not be congested and thus, the prosumers do not have market power to manipulate the prices, and the bidding converges to a unique GNE.

The two-bus example above shows some nice properties of GNE(s) of game $\mathcal{G}$ in \eqref{eq:sharing-game}, such as existence and optimality in terms of \eqref{eq:central}, as well as uniqueness if no congestion occurs at the optimal solution of  \eqref{eq:central}. However, these properties cannot be readily extended to general networks, as counter-examples are identified with the following three-prosumer model.

\begin{figure}[h]
\centerline{\includegraphics[width=0.85\columnwidth]{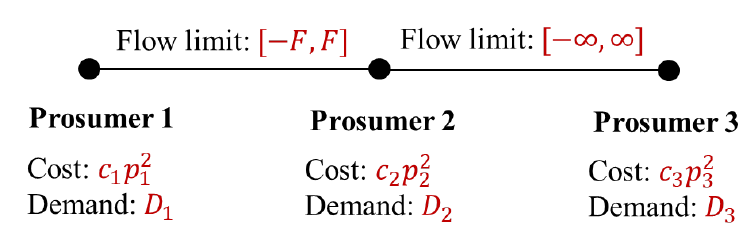}}
\setlength{\abovecaptionskip}{0pt}
\caption{Settings of Example 2 with three prosumers.}
\label{fig:example2}
\end{figure}

\textbf{Example 2:} Three prosumers are connected as shown in Fig. \ref{fig:example2}. Set $a=1$, $c_1=c_2=c_3=c$, $d_1=d_2=d_3=0$, $E_i^0=D_i,\forall i=1,2,3$, and flow limit $F$ for the line connecting prosumers 1 and 2.

The GNE(s) for the three-prosumer example can be analyzed in a way similar to the two-prosumer example, although details are more involved and thus elaborated in Appendix \ref{apen-example}. In summary, if both lower and upper line flow constraints are inactive at the unique optimal solution of \eqref{eq:central}, then game $\mathcal{G}$ in \eqref{eq:sharing-game} may either have no GNE or a unique GNE $(\bar p, \bar b)$ where $\bar p$ is optimal for \eqref{eq:central}. If one side of the line flow constraints is binding at the unique optimal solution of \eqref{eq:central}, then game $\mathcal{G}$, in general, may have no GNE or uncountably many GNEs; in the latter case, some of the GNEs have the same power profile $\bar p$ which is optimal for \eqref{eq:central}, while others do not. 
There are also special cases in which the game has a unique GNE: one example is that a line flow constraint is barely reached but not binding; another example is that a line flow constraint is binding and network parameters are just set on the boundary between no-GNE and multi-GNE cases. These special cases are rare in practice and hence not discussed in detail.

The two examples above show that for game $\mathcal{G}$ \eqref{eq:sharing-game} in general, there is no guarantee for existence, uniqueness, or optimality of GNE, which exerts challenges on market operation. To tackle this problem, we develop an improved energy sharing mechanism with price regulation in Section \ref{sec-4}.


\section{Improved energy Sharing Mechanism}
\label{sec-4}
We have shown that when network constraint is taken into account, the situation of GNE of the energy sharing game $\mathcal{G}$ \eqref{eq:sharing-game} is quite unpredictable, which is challenging for market operation. 
In this section, a price regulation policy is proposed to derive an improved energy sharing mechanism. We prove existence and uniqueness of GNE for this improved mechanism and show that a Pareto improvement can be achieved. Tendency of the GNE with increasing prosumers and the structure of the energy sharing price are also revealed.

\subsection{Price regulation policy}
\label{sec-4.1}
We regulate energy sharing price of prosumer $i \in\mathcal{I}$ as:
\begin{align}
\label{eq:priceregulate}
\lambda_i^r = \begin{cases}
\max \left\{ \lambda_i(b), ~2c_ip_i+d_i-\frac{q_i(b)}{a(I-1)} \right\}, & q_i(b) \ge 0 \\
\min \left\{ \lambda_i(b), ~2c_ip_i+d_i-\frac{q_i(b)}{a(I-1)} \right\}, & q_i(b) < 0
\end{cases}
\end{align}
where $\lambda_i(b)$ is the $i$-th element of the optimal solution of problem \eqref{eq:new-rule} given $b$. The regulated price \eqref{eq:priceregulate} leads to an \emph{improved energy sharing mechanism} described by the following decision-making problem for every prosumer $i \in \mathcal{I}$:
\bsq
\label{eq:modified-obj}
\begin{eqnarray}
\mathop {\min} \limits_{p_i,b_i} &~& c_ip_i^2+d_ip_i+u_i(b,p_i) \label{eq:modified-obj.1}\\
\mbox{s.t.} &~& p_i+\underbrace{(-a\lambda_i(b)+b_i)}_{q_i(b)}=D_i \label{eq:modified-obj.2}
\end{eqnarray}
\esq
where

\begin{eqnarray}
u_i(b,p_i)  = \max \bigg\{\lambda _i(b)\cdot q_i(b), \left(2c_i p_i + d_i - \frac{q_i}{a(I - 1)}\right)\cdot q_i(b) \bigg\} \nonumber 
\end{eqnarray}

\textbf{Remark}: An implication of the price regulation policy \eqref{eq:priceregulate} is that it restricts the price privilege granted to every prosumer $i \in \mathcal{I}$ compared to its marginal production adjustment cost. Formally, the price privilege of prosumer $i$ is $2c_i p_i+d_i-\lambda_i^r$ if it is a buyer ($q_i = D_i - p_i \geq 0$) and $\lambda_i^r - (2c_i p_i+d_i)$ if it is a seller ($q_i <0$). 
With price regulation, the maximum price privilege of prosumer $i$ is restricted to $|q_i|/{a(I-1)}$, which can be regarded as prosumer $i$'s \emph{degree of participation in sharing} characterized by its sharing amount $|q_i|$, market sensitivity factor $a$, and the total number $I$ of prosumers. 

In the following, we show that the improved energy sharing mechanism (with price regulation) has desired properties.

\begin{theorem}
\label{Thm:prop-2}
The improved energy sharing mechanism \eqref{eq:modified-obj} attains a \emph{unique} GNE $(\bar p ,\bar b)$ with regulated price $\bar \lambda^r$, where $\bar{p}$ is the unique optimal solution of \eqref{eq:central} and 
\bsq\label{eq:cond:improved-GNE}
\begin{align}
    \bar \lambda_i^r &= \lambda_i = 2c_i \bar p_i+d_i -\frac{D_i-\bar p_i}{a(I-1)}, \quad \forall i\in \mathcal{I}  \label{eq:cond:improved-GNE.1} \\
        \bar b_i & = D_i-\bar p_i + a \bar \lambda_i^r,\quad \forall i\in \mathcal{I} 
\end{align}
\esq
\end{theorem}
The proof of Theorem \ref{Thm:prop-2} is in Appendix \ref{apen-5}. It shows that by including a price regulation policy, a unique GNE of the improved energy sharing mechanism always exists, which circumvents the no-GNE and multi-GNE issues encountered by the original game \eqref{eq:sharing-game}. 
Besides, the improved mechanism provides a simpler model for GNE computation by establishing equivalence between GNE and the optimal solution of \eqref{eq:central}.

\subsection{Incentive for prosumers}
\label{sec-4.2}
The following proposition shows that every prosumer is incentivized to participate in the improved energy sharing market by incurring a cost not exceeding self-sufficiency.

\begin{proposition}
\label{Thm:prop-3} 
For every prosumer $i\in \mathcal{I}$, denote its modified objective \eqref{eq:modified-obj.1} as $\tilde \Gamma_i(p,b)$, and recall $J_i(p_i)=c_ip_i^2 +d_i p_i$. The following inequality holds at the unique GNE $(\bar p,\bar b)$ of the improved energy sharing game \eqref{eq:modified-obj}:
\bq
\label{eq:pareto}
 \tilde \Gamma_i (\bar p,\bar b) \le J_i(D_i), ~\forall i\in \mathcal{I} 
\eq
\end{proposition}

Proposition \ref{Thm:prop-3} is proved in Appendix \ref{apen-4}. It shows that prosumer $i \in \mathcal{I}$'s cost in the energy sharing market is no more than its cost under self-sufficiency. Essentially, the improved sharing mechanism achieves a Pareto improvement over self-sufficiency and thus leaves \emph{no} prosumer worse off. Thus, prosumer $i \in \mathcal{I}$ always has the incentive to implement $q_i$ as requied by the platform. A contract can be signed beforehand that prosumers who are not responding accordingly will face a high penalty or be barred from the market. If one prosumer has an equipment failure, we can exclude that prosumer from the energy sharing market and let the other prosumers bid again. We will show later that it only takes a few seconds for the proposed bidding algorithm to converge to a new equilibrium.

\subsection{Tendency with a growing number of prosumers}
\label{sec-4.3}
Welfare loss occurs at the GNE of the improved energy sharing game compared to the social optimum. 
Our next proposition quantifies and bounds this loss by price of anarchy.

\begin{definition} (Price of Anarchy, PoA \cite{johari2011parameterized}) 
Let $J(p):=\sum_{i \in \mathcal{I}} J_i(p_i)$ be the measure of market efficiency. \emph{Price of Anarchy (PoA)} is defined as the ratio between the values of $J(p)$ at the worst equilibrium and the social optimum.
\end{definition}

\begin{proposition}
\label{Thm:prop-5} 
Given prosumer number $I$, let $(\bar p(I),\bar b(I))$ be the unique GNE of the improved sharing game \eqref{eq:modified-obj} and $\tilde p(I)$ be the social optimum for \eqref{eq:AGG}. 
We assume there is a \emph{uniform} upper bound $C_1 \geq \max\{(D_i - \tilde p_i)^2,~ (D_i - \bar p_i)^2\}$, as well as a \emph{uniform} lower bound $0< C_2 \leq J_i(\tilde p_i)$, for all number $I$ and $i\in \mathcal{I}$.
The PoA of the improved energy sharing game satisfies:
\begin{align}
    1 \le \mbox{PoA}(I) :=\frac{J(\bar p(I))}{J(\tilde p(I))} \le  1+\frac{C_1}{2 a(I-1)C_2} \label{eq:RWLbounds}
\end{align}
\end{proposition}

Since the improved energy sharing game \eqref{eq:modified-obj} has a unique GNE, which is also the ``worst'' equilibrium, its PoA equals the ratio between $J(\bar p)$ and $J(\tilde p)$.
The proof of Proposition \ref{Thm:prop-5} is in Appendix \ref{apen-6}. It implies $\lim_{I\rightarrow \infty} \mbox{PoA}(I)=1$, i.e., the total disutility of prosumers at the GNE of the improved energy sharing game approaches the social optimum as the number of prosumers increases. In other words, the improved mechanism can still achieve social optimum in a large market as the original mechanism (as proved in Proposition \ref{Thm:prop-0}).

\subsection{Structure of the energy sharing price}
\label{sec-4.4}
The energy sharing price in our improved mechanism has an elegant structure as presented by the following proposition.
 
\begin{proposition}
\label{Thm:prop-4}
At the unique GNE $(\bar p,\bar b)$ of the improved energy sharing game \eqref{eq:modified-obj}, the energy sharing price is:
\begin{align}
\bar \lambda_i^{r}  = \lambda_i(\bar b) =-\bar \kappa-\sum \limits_{l=1}^L \pi_{il}\bar \tau_{l}^{-}+\sum \limits_{l=1}^L \pi_{il}\bar \tau_{l}^{+},~\forall i\in \mathcal{I} \label{eq:price-structure}
\end{align}
where $(\bar \kappa, \bar \tau^{\pm})$ is any dual optimal solution of problem \eqref{eq:central}.
\end{proposition}

The proof of Proposition \ref{Thm:prop-4} is in Appendix \ref{apen-7}. 
The price $\bar \lambda_i^{r}$ in \eqref{eq:price-structure} is composed of $-\bar \kappa$, the price for network-wide energy balancing, and $-\sum_{l=1}^L \pi_{il}\bar \tau_{l}^{-}+\sum_{l=1}^L \pi_{il}\bar \tau_{l}^{+}$, the price of line congestion incident to prosumer $i \in \mathcal{I}$.
This structure is similar to the classic locational marginal price (LMP) \cite{bose2019some}.

The price structure \eqref{eq:price-structure} implies that no subsidy is required to run the proposed energy sharing market, because the net sharing payment of all the prosumers at GNE is nonnegative as calculated below (where $\bar \lambda := \bar \lambda^r = \lambda (\bar b)$):
\begin{align}
~ & \sum\limits_{i=1}^I \bar \lambda_i^{r}(-a\lambda_i(\bar b)+\bar b_i) = \sum\limits_{i=1}^I \bar \lambda_i(-a\bar \lambda_i+\bar b_i) \nonumber\\
=~ & \sum \limits_{i=1}^I (-\bar \kappa-\sum \limits_{l=1}^L \pi_{il}\bar \tau_l^{-}+\sum \limits_{l=1}^L \pi_{il} \bar \tau_l^{+})(-a\bar \lambda_i+ \bar b_i) \nonumber\\
 = ~ & \sum \limits_{l=1}^L [-\sum \limits_{i=1}^I \pi_{il}(-a\bar \lambda_i+\bar b_i)\bar \tau_l^{-}+\sum \limits_{i=1}^I \pi_{il} (-a\bar \lambda_i+\bar b_i)\bar \tau_l^{+}] \nonumber\\
=~ & \sum \limits_{l=1}^L \left(\tilde F_l\bar \tau_l^{-}+F_l\bar \tau_l^{+}\right) \ge 0 \nonumber
\end{align}
The second and third equalities are due to the energy balance and the complementary and slackness condition for line congestion, respectively. If no congestion occurs at GNE, then $\bar \tau^{\pm}=0$, the net sharing payment is zero, and the market is economically self-balanced.
In the presence of congestion, by energy sharing, not only every prosumer is better off than self-sufficiency, but also the market platform receives a revenue. Note that the measure of market efficiency (social welfare) in terms of cost has already taken into account the merchandising surplus, which is the sum of prosumer costs minus the merchandising surplus of the platform, i.e., $\sum_i (J_i(p_i)+\lambda_iq_i)-\sum_i \lambda_iq_i=\sum_i J_i(p_i)=J(p)$. The social welfare loss is thus defined as $J(\bar p)-J(\tilde p)$, where $\bar p$ and $\tilde p$ are the net production adjustment under GNE and social optimum, respectively. This is consistent with the definition of deadweight loss \cite{hirshleifer2005price} in economics, which includes both the supplier/consumer surplus and the government tax revenue in social welfare calculation.

\section{Bidding Algorithm to Achieve GNE}
\label{sec-5}
We have proved a set of desired properties of the GNE of the improved energy sharing mechanism. How to reach this GNE is also a crucial issue. This section presents a practical bidding algorithm and provides guidance on parameter selection to guarantee convergence of the bidding algorithm to the GNE. \begin{algorithm}[t]
	\caption{Bidding Algorithm for Energy Sharing}
	\KwIn{$I$, $a$, $c_i$, $d_i$, $D_i$ to the smart meter of every prosumer $i \in \mathcal{I}$; tolerance $\epsilon>0$.}
	\KwOut{energy sharing results $p^*,b^*,\lambda^*$.}
	\textbf{Initialization:} $b^1=\lambda^1=0$, $k=0$\; 
	\Repeat{$\|b^{k+1}-b^k\| \le \epsilon$}{
		iteration $k++$
		
		\textbf{platform update:}
		\begin{align}
		\label{eq:platform-update2}
		    \lambda^{k+1}=\mbox{argmin}_{\lambda}~ & \sum \limits_{i=1}^I \lambda_i^2+\sum \limits_{i=1}^I (\lambda_i-\lambda_i^{k})^2 \nonumber\\
		    \mbox{s.t.}~ & \sum \limits_{i=1}^I (-a\lambda_i+b_i^k)=0 \\
		    ~ & -\tilde F_l \le \sum \limits_{i=1}^I \pi_{il} (-a\lambda_i+b_i^k) \le F_l,\forall l \in \mathcal{L} \nonumber
		\end{align}
		
		\textbf{prosumer update:}
		
		\For{each $i \in \mathcal{I}$}
		{ 
		\begin{align}
             p_i^{k+1}~ & \mbox{is set according to \eqref{eq:prosumer_update}} \nonumber\\
		    b_i^{k+1}~ & =D_i-p_i^{k+1}+a\lambda_i^{k+1} \nonumber
		\end{align}
		}
	}
\end{algorithm}

The bidding algorithm, consisting of \emph{platform update} and \emph{prosumer update}, is elaborated in \textbf{Algorithm 1}. Specifically:

1) \textbf{Platform Update}. For the platform, we hope to be as fair as possible by minimizing the variance of prices across prosumers; and meanwhile, we want to keep the prices stable by minimizing their deviations from the prices in the last iteration. Therefore, an additional term $\sum \nolimits_{i=1}^I (\lambda_i-\lambda_i^k)^2$ is included in the objective function \eqref{eq:new-rule.1} to avoid severe fluctuation of market prices across the bidding process. Given prosumer bids, the energy sharing price is solved from \eqref{eq:platform-update2}.

2) \textbf{Prosumer Update}. Prosumers are price-makers, and they will estimate the impact of their current bids on the energy sharing price in the next iteration. When deciding on its current bid, prosumer $i$ solves problem \eqref{eq:modified-obj} taking into account the change of $\lambda$ incurred by $b$ via \eqref{eq:new-rule}. In particular, in the $k^{\mbox{th}}$ iteration, each prosumer $i \in \mathcal{I}$ utilizes up-to-date price $\lambda_i^{k+1}$ to replace $\bar \lambda_i^r$ in \eqref{eq:cond:improved-GNE.1} and estimate its optimal strategy:
\begin{align}
    p_i^{k+1}=\frac{a(I-1)\lambda_i^{k+1}-a(I-1)d_i+D_i}{2a(I-1)c_i+1} \label{eq:prosumer_update}
\end{align}

The market sensitivity $a$ is the key factor that influences the convergence of the bidding algorithm. Since $q_i=-a\lambda_i+b_i$, if $a$ is too small, a subtle change in $\lambda_i$ will result in a strong reaction in $b_i$. This may lead to significant oscillation and failure to converge to a stable market equilibrium. We provide guidance for the policy makers to choose the market sensitivity $a$ by Condition A2 below, which specifies a range of $a$ within which we can prove convergence of \textbf{Algorithm 1}.


\textbf{A2:} $a \ge \frac{I-2}{2(I-1)} {\mbox{max}_{i \in \mathcal{I}}} \{1/c_i\}$.

\begin{theorem}
\label{Thm:prop-7}
When A2 holds, Algorithm 1 converges to the unique GNE of the improved energy sharing game \eqref{eq:modified-obj}.
\end{theorem}

\begin{figure}[h]
\centerline{\includegraphics[width=1.0\columnwidth]{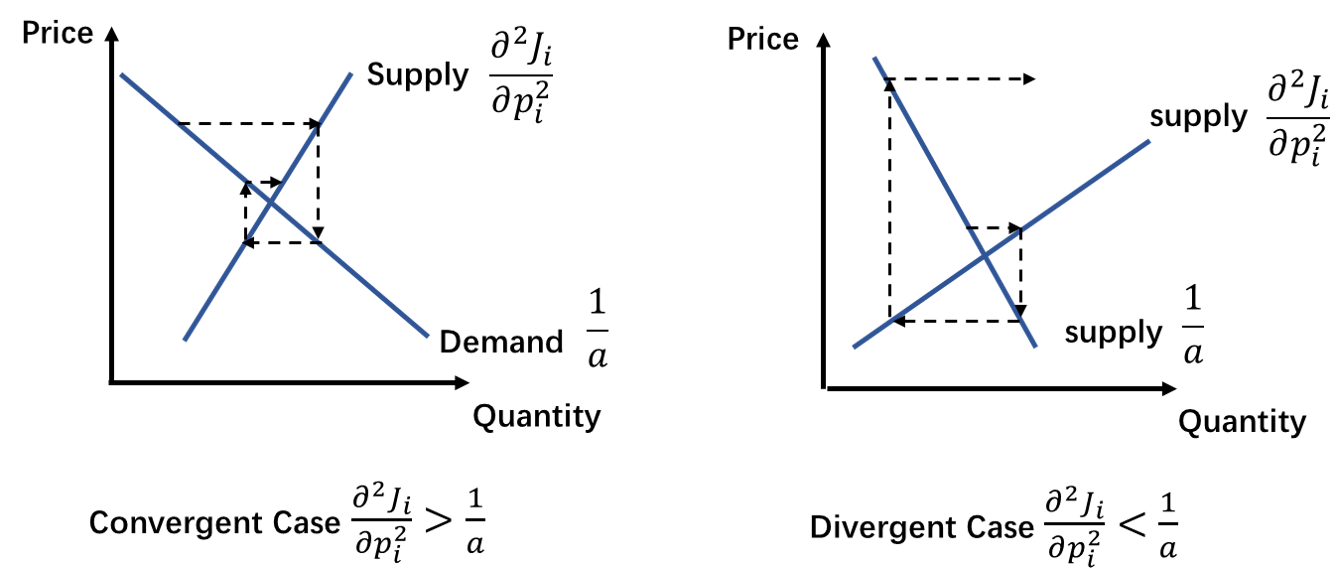}}
\caption{Economic interpretation of Condition A2 based on cobweb model. The slope of supply curve is $\partial^2 J_i/\partial p_i^2$ and the slope of the demand curve is $-1/a$. In a large market, when $\partial^2 J_i/\partial p_i^2>1/a$, the bidding algorithm converges as in the left hand side of the figure; when $\partial^2 J_i/\partial p_i^2<1/a$, the bidding algorithm diverges as in the right hand side of the figure.}
\label{fig:cobweb}
\end{figure}

The proof of Theorem \ref{Thm:prop-7} is in Appendix \ref{apen-8}. Condition A2 provides a lower bound for $a$, but to protect privacy, we do not need to know the exact value of the lower bound. For example, we can let each prosumer $i \in \mathcal{I}$ to report a $c_i-\epsilon_i$ with a random noise $\epsilon_i>0$, and set the parameter   according to \eqref{eq:privacy} so that A2 is satisfied.
\begin{align} \label{eq:privacy}
    a \ge \frac{I-2}{2(I-1)}\max_{i \in \mathcal{I}} \{1/(c_i-\epsilon_i)\} \ge \frac{I-2}{2(I-1)}\max_{i \in \mathcal{I}} \{1/c_i\}
\end{align}

We try to explain the economic intuition of Condition A2 using the well-known cobweb model in economics as in Fig. \ref{fig:cobweb}. The condition is equivalent to $1/a \le \frac{2(I-1)}{I-2} {\mbox{min}_{i \in \mathcal{I}}} \{c_i\}=\frac{I-1}{I-2} \mbox{min}_{i \in \mathcal{I}} \{\partial^2 J_i / \partial p_i^2\}$. Note that $\partial J_i /\partial p_i$ is the marginal disutility of prosumer $i \in \mathcal{I}$ and can be regarded as the supply curve of production adjustment, where $\partial^2 J_i/\partial p_i^2$ is the slope of this supply curve. The demand for production adjustment comes from the energy sharing market, and since $q_i=-a\lambda_i+b_i$, the slope of demand curve is $-1/a$. The term $\frac{I-1}{I-2}$ in Condition A2 is caused by the ability of individual prosumer to influence the market price, and when $I \to \infty$ this term goes to 1. In a large market, as in Fig. \ref{fig:cobweb}, the bidding algorithm converges if and only if $\partial^2 J_i/\partial p_i^2>1/a,\forall i \in \mathcal{I}$.  Furthermore, when $a$ is small ($1/a$ is large), the sharing price responds more quickly to the changing bids, and therefore, the algorithm can reach the equilibrium faster. When $I \to \infty$, the prosumer update rule \eqref{eq:prosumer_update} becomes $2c_ip_i^{k+1}+d_i=\lambda_i^{k+1}$. 
At this time, the bidding algorithm (Algorithm 1) has the same form as the alternating direction method of multipliers (ADMM) \cite{boyd2011distributed} to solve problem \eqref{eq:AGG}. This shows that with a growing number of prosumers, both the equilibrium (Proposition \ref{Thm:prop-5}) and the bidding process of the improved energy sharing market approach those under centralized operation (social optimum).

\textbf{Remark on market sensitivity parameter $a$.}  In practice, the $a_i$ is a private information of prosumer $i \in \mathcal{I}$ and may be hard to obtain even by the prosumer itself. Here, we conjecture that prosumers participating in the same energy sharing market usually have similar sensitivity, so for simplicity of analysis, a uniform $a$ is used as a parameter in the set rule for sharing market clearing. We set the parameter based on condition A2 to ensure that the bidding algorithm (Algorithm 1) converges to the unique GNE of the improved energy sharing game. Moreover, as proved in Proposition \ref{Thm:prop-3}, prosumers always have the incentive to join the energy sharing market as long as $a>0$, since they will not be worse-off. Therefore, even though this uniform $a$ does not match each prosumer’s individual price sensitivity accurately, it is still willing to take part in energy sharing. Setting the parameter $a$ by A2 is also reasonable.

\textbf{Remark on feasibility}: First, in the $k$-th iteration, given all prosumers’ bids $b_i^k,\forall i \in \mathcal{I}$, the platform solves problem \eqref{eq:platform-update2} to update the prices. We can easily construct a feasible solution by letting $\lambda_i=b_i^k/a,\forall i \in \mathcal{I}$. Therefore, problem \eqref{eq:platform-update2} is always feasible. Second, as proved in Theorems \ref{Thm:prop-2}-\ref{Thm:prop-7}, the bidding algorithm will converge to the optimal solution of problem \eqref{eq:central}. Problems \eqref{eq:AGG} and \eqref{eq:central} have the same feasible set, and due to assumption A1, problem \eqref{eq:central} is also feasible.


\section{Possible Extensions} \label{sec:extension}
The main goal of this paper is to study the strategic behavior of individual prosumers in a sharing market when network constraints are considered. As a first step, we study a simplified situation (power balance) that captures essential features of the sharing scheme without obscuring its fundamental properties. We discuss some possible extensions below.

1) \textbf{Arbitrary convex cost functions \& capacity constraints.} In this paper, we focus on the quadratic cost structure since many of the costs/utilities in power networks can be represented as quadratic functions, such as the cost for generator \cite{wei2014robust}, the utility for a consumer \cite{samadi2012advanced}, the cost for curtailment \cite{wu2014robust}, and so on. But it is worth mentioning that in our previous work \cite{chen2020approaching}, several properties similar to those in this paper have been proved for an energy sharing market with arbitrary convex cost functions and capacity constraints, but \emph{without consideration of networks}. We believe that many properties of the proposed networked energy sharing mechanism can also be generalized, although that will require more complicated theoretical analysis. This is among our undergoing works.

2) \textbf{Financial incentives in demand response.} For notation conciseness, in this paper, we did not model the financial incentives/penalties in demand response explicitly. In the following, we show how the proposed model be extended to incorporate that. Let $\Delta_i$ be the amount of energy reduction requirement that is not met, and the associated penalty is $\tilde c_i \Delta_i^2$. Then the prosumer’s problem can be formulated as
\begin{align}\label{eq:penalty}
    \min_{p_i,\Delta_i,b_i}~ & c_ip_i^2+d_ip_i + \tilde c_i \Delta_i^2 + \lambda_i(-a\lambda_i+b_i) \nonumber\\
    \mbox{s.t.}~ & p_i+\Delta_i+(-a\lambda_i+b_i)=D_i
\end{align}
We can prove that the optimal solution of \eqref{eq:penalty} is equivalent to that of \eqref{eq:penalty-eq}.
\begin{align}\label{eq:penalty-eq}
    \min_{\tilde p_i,b_i}~ & \left(\frac{1}{1/(2c_i)+1/(2\tilde c_i)}\right)\left(\frac{1}{2}\tilde p_i^2+\frac{d_i}{2c_i} \tilde p_i\right) + \lambda_i (-a\lambda_i+b_i) \nonumber\\
    \mbox{s.t.}~ & \tilde p_i+(-a\lambda_i+b_i)=D_i 
\end{align}
with
$$p_i=\frac{\tilde p_i-\frac{d_i}{2\tilde c_i}}{1+\frac{c_i}{\tilde c_i}},~ \Delta_i=\frac{\tilde p_i + \frac{d_i}{2c_i}}{1+\frac{\tilde c_i}{c_i}}$$
Problem \eqref{eq:penalty-eq} follows the same form as the problem we studied in the paper, so we can use \eqref{eq:penalty-eq} to analyze and recover $p_i$ and $\Delta_i$ afterwards.

3) \textbf{Application scenarios}. This paper targets at a distribution system with flexible prosumers. The typical distribution systems, such as the IEEE 33-bus and IEEE 69-bus systems, are tested. But it is worth mentioning that the proposed model and method are quite general and can be also used for applications in transmission systems. For example, the prosumer can be replaced with an industry area with both power plants and the demand for electricity.

4) \textbf{AC power flow}. In this paper, the DC power flow is used for simplicity. The voltage variables are eliminated by assuming that their magnitudes are near 1.0 per unit, so there is no voltage constraint \cite{stott2009dc}. To be more accurate, we can replace the market clearing constraints \eqref{eq:new-rule.2}-\eqref{eq:new-rule.3} by AC power flow models to incorporate voltage constraints. However, this will greatly complicate the analysis due to its nonconvexity. This is also one of our undergoing works.

\section{Case Studies}
\label{sec-6}
Numerical experiments were conducted to validate the propositions and theorems in this paper.
\subsection{A simple case with two prosumers}
\label{sec-6.1}
\begin{figure}[h]
\centerline{\includegraphics[width=1.0\columnwidth]{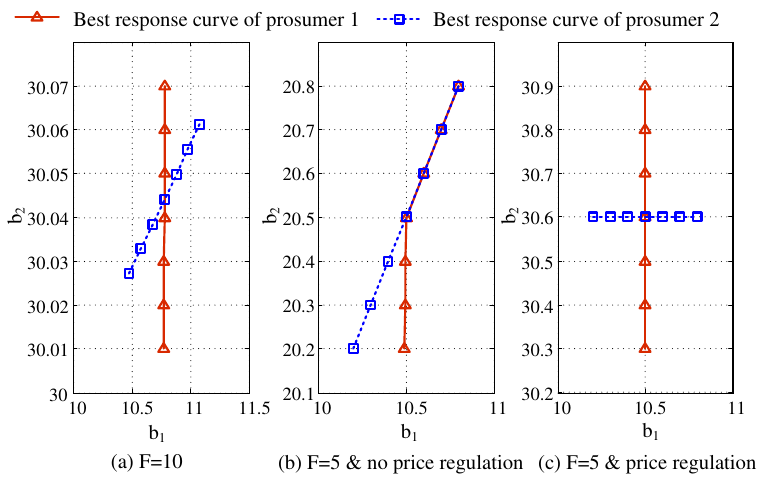}}
\setlength{\abovecaptionskip}{0pt}
\caption{Best response curves of two prosumers.}
\label{fig:BRcurve}
\end{figure}

We first test a simple case with two prosumers connected by a line. The parameters for the two prosumers are: $c_1 = 0.003 \; \mbox{\$/kW}^2$, $d_1 = 0.42 \; \mbox{\$/kW}$, $D_1 = 100 \; \mbox{kW}$; and $c_2 = 0.006 \; \mbox{\$/kW}^2$, $d_2 = 0.72 \; \mbox{\$/kW}$, $D_2 = 200 \; \mbox{kW}$. The market sensitivity is chosen as $a=10$. We change the value of flow limit $F$ and test the cases with/without price regulation; the results are illustrated in Fig. \ref{fig:BRcurve}.

When $F=10$, the line is not congested at the optimal point of \eqref{eq:central}, and as shown in Fig. \ref{fig:BRcurve}(a) there is a unique GNE $(p_1,b_1)=(109.6,10.78)\; \mbox{kW}$, $(p_2,b_2)=(190.4,30.04) \; \mbox{kW}$. When $F=5$, the line is congested at the optimal point of \eqref{eq:central}. Without price regulation, as shown in Fig. \ref{fig:BRcurve}(b), there are multiple GNEs. With price regulation, as shown in Fig. \ref{fig:BRcurve}(c), there is a unique GNE $(p_1,b_1)=(105.0,10.50)\; \mbox{kW}$, $(p_2,b_2)=(195.0,30.60) \; \mbox{kW}$, which validates Theorem \ref{Thm:prop-2}. A comparison of self-sufficiency, energy sharing, and social optimum is given in TABLE \ref{tab:comparison} for $F=5$, with price regulation and the same parameters above. To be specific, self-sufficiency refers to the case in which each prosumer adjusts its own net production to meet the required energy purchase reduction. Under the energy sharing scheme, prosumers can trade with each other in the proposed improved energy sharing market based on Algorithm 1. Social optimum is obtained by solving problem \eqref{eq:AGG}. Total cost is the sum of two prosumers’ costs, and $J(p)=\sum_i J_i(p_i)$ is the measure of market efficiency.
\begin{table}[h]
	\renewcommand{\arraystretch}{1.3}
	\centering
	\caption{Comparison of three schemes}
	\label{tab:comparison}
	\begin{tabular}{ccccc}
		\hline 
		Paradigm & Self-Sufficiency & Energy Sharing & Social Optimum  \\
		\hline
		$p_1,p_2$ (kW) & 100, 200 & 105, 195 & 105, 195 \\
		Cost (\$) & 72.0, 384.0 & 69.4, 381.4 & 77.2, 368.6 \\ 
		Total (\$)& 456.0 & 450.8 &  445.7 \\
		$J(p)$ (\$) & 456.0 &  445.7 & 445.7 \\
		\hline
	\end{tabular}
\end{table}

\begin{figure}[h]
\centerline{\includegraphics[width=1.0\columnwidth]{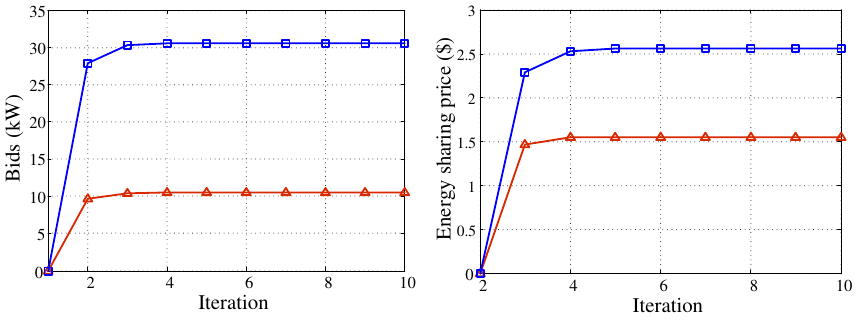}}
\setlength{\abovecaptionskip}{0pt}
\caption{Bids and energy sharing prices during iterations of Algorithm 1.}
\label{fig:convergence}
\end{figure}

Several insights can be derived from TABLE \ref{tab:comparison}. Compared with self-sufficiency, each prosumer is better-off by energy sharing. Take prosumer 1 as an example: Under self-sufficiency, it increases it net production by 100kW to meet the energy reduction requirement $D_1$, resulting in a cost of \$72 ($=0.003\times 100^2+0.42 \times 100$). Under energy sharing, prosumer 1 increases its net production by 105kW, and sells 5kW to prosumer 2 at an energy sharing price of 1.55\$/kW. So the cost of prosumer 1 reduces to \$69.43 ($=0.003\times 105^2+0.42 \times 105-1.55\times 5$). Similarly, the cost of prosumer 2 decreases from \$384 ($=0.006\times 200^2+0.72 \times 200$) to \$381.4 ($=0.006 \times 195^2+0.72 \times 195+2.56 \times 5$). By letting prosumer 1 with lower cost to produce more and sell energy to prosumer 2, a Pareto improvement can be reached, which verifies Proposition \ref{Thm:prop-3}. Although the GNE of energy sharing minimizes \eqref{eq:central.1} and the centralized operation minimizes a different objective \eqref{eq:AGG.1}, due to the binding line limit, the two mechanisms in our example attain the same solution with $J(p)=445.7$, which is lower than $J(p)=456.0$ under self-sufficiency.
We apply Algorithm 1 to achieve the GNE, and the changes of bids and energy sharing prices are recorded in Fig. \ref{fig:convergence}. The algorithm converges to the GNE after 4 iterations.

We further illustrate how the market roles of prosumers are endogenously given in TABLE \ref{tab:marketrole}. When prosumer 2 shares with prosumer 1, it is a buyer; but when it shares with prosumer 3, it becomes a seller.

\begin{table}[h]
	\renewcommand{\arraystretch}{1.3}
	\centering
	\caption{Prosumers’ market roles under different settings}
	\label{tab:marketrole}
	\begin{tabular}{cccc}
		\hline 
        & Prosumer-1 & Prosumer-2 & Prosumer-3 \\
        \hline
        $c_i,d_i,D_i$ & (0.003, 0.42, 100) & 	(0.006, 0.72, 200) & (0.008, 0.72, 200) \\
        1, 2 share &  Sell 5 & Buy 5 & / \\
        1, 3 share & Sell 5 & / & Buy 5\\
        2, 3 share & / & Sell 5 & Buy 5\\
		\hline
	\end{tabular}
\end{table}

\subsection{IEEE 38-bus test system}
\label{sec-6.2}
\begin{figure}[h]
\centerline{\includegraphics[width=0.7\columnwidth]{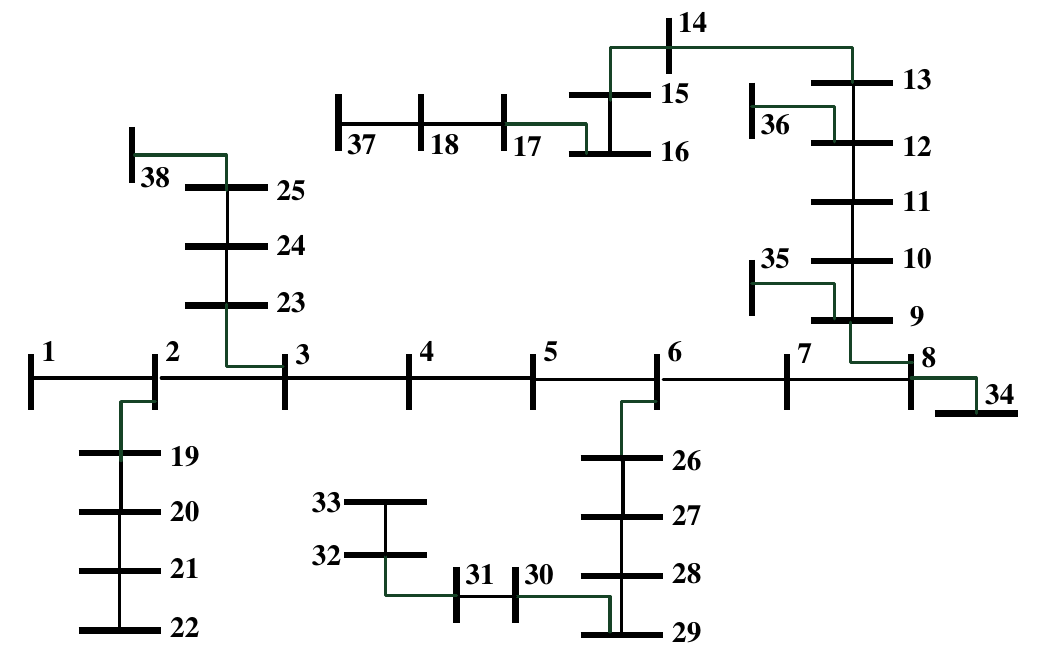}}
\setlength{\abovecaptionskip}{0pt}
\caption{The topology of the 38-bus test system.}
\label{fig:topology}
\end{figure}

\begin{figure}[h]
\centerline{\includegraphics[width=0.9\columnwidth]{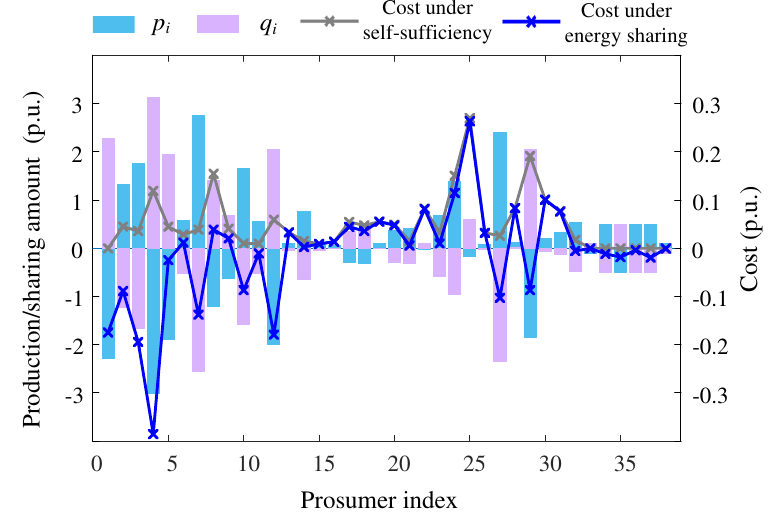}}
\setlength{\abovecaptionskip}{0pt}
\caption{Energy sharing market outcome in IEEE 38-bus microgrid.}
\label{fig:result-case38}
\end{figure}

\begin{figure}[h]
\centerline{\includegraphics[width=1.0\columnwidth]{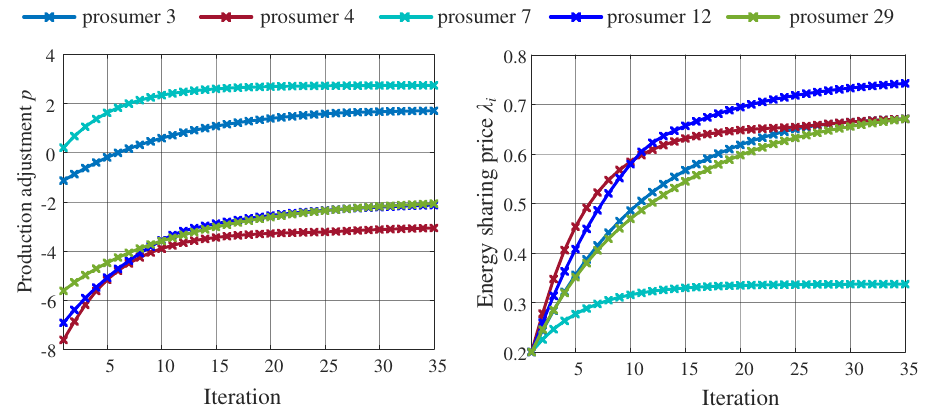}}
\setlength{\abovecaptionskip}{0pt}
\caption{Production adjustments and energy sharing prices during iterations.}
\label{fig:case38-bidding}
\end{figure}

We test a modified IEEE 38-bus microgrid model \cite{singh2009multiobjective} to validate scalability of the proposed mechanism. The p.u. unit for energy is 100kWh, for cost is \$50, and for price is 0.5\$/kWh. The test system is shown in Fig. \ref{fig:topology}. The production $p_i$, sharing quantity $q_i$, and cost $\tilde \Gamma_i(p,b)$ at the GNE of the improved energy sharing mechanism, as well as the cost $J_i(D_i)$ under self-sufficiency, are plotted in Fig. \ref{fig:result-case38}. Taking prosumer 4 for example, its required energy reduction is $D_4=0.12$ (p.u.), which calls for an increase of $0.12$ (p.u.) in production under self-sufficiency. However, under the sharing mechanism, it instead reduces its production by $3.00$ (p.u.) and buys $3.12$ (p.u.) from the sharing market, which makes a profit of $0.38$ (p.u.). The grey curve in the figure refers to the cost under self-sufficiency while the blue curve refers to that under energy sharing. We can find that the blue curve is always lower than the grey curve, meaning that no prosumer gets worse-off. Many prosumers (e.g. prosumers 1-12, 27, 29) can even earn profits from sharing.
The average profit improvement of all the prosumers is 0.0629 (p.u.). In average, the prosumers are turning from paying (0.0511 p.u.) to earning (-0.0118 p.u.), with a cost reduction of 123.1\%. 
This shows the great potential of our proposed mechanism in improving social welfare. We further test the performance of the proposed bidding algorithm on the 38-bus system. We pick up the prosumers who benefit the most from energy sharing compared to self-sufficiency (i.e. prosumers 3, 4, 7, 12, 29) and record their changes of production adjustments and energy sharing prices during iterations in Fig. \ref{fig:case38-bidding}. All of them converge to the GNE quickly in 35 iterations.

\begin{figure}[h]
\centerline{\includegraphics[width=0.7\columnwidth]{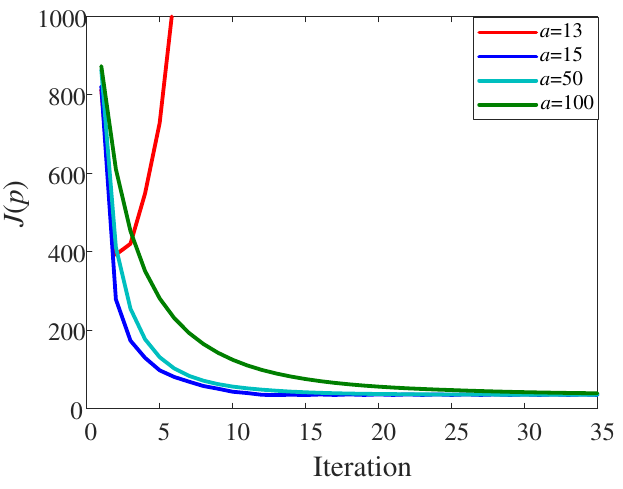}}
\setlength{\abovecaptionskip}{0pt}
\caption{The change of $J(p)$ during iterations with different $a$.}
\label{fig:figureR1}
\end{figure}

We then test the performance of the bidding algorithm under different $a$. The change of $J(p)$ is shown in Fig. \ref{fig:figureR1}. For the IEEE-38 bus system, Condition A2 gives $a \ge 33.04$. We can find that when $a$ is too small ($a=13$), the algorithm diverges. Condition A2 is a sufficient but not necessary condition. Even if $a$ is chosen as 15 that violates A2, the algorithm still converges. The larger the $a$, the slower the algorithm converges. 

\begin{figure}[h]
\centerline{\includegraphics[width=0.75\columnwidth]{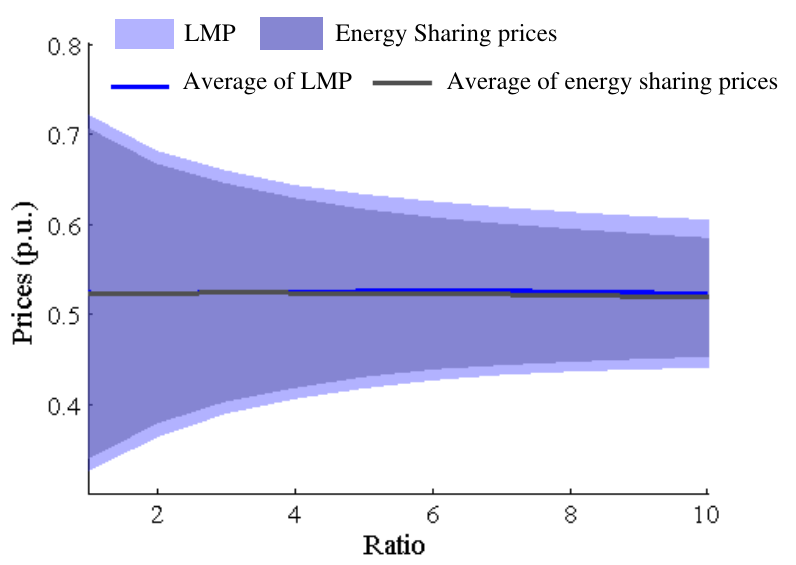}}
\setlength{\abovecaptionskip}{0pt}
\caption{LMP and energy sharing prices under different flow limits.}
\label{fig:prices}
\end{figure}

We simultaneously tune the flow limits of all the lines from 1 to 10 times their original values. 
Fig. \ref{fig:prices} shows over different flow limits the average and the max-min span of nodal prices at the social optimum (where nodal prices are LMPs) and at the GNE of the improved energy sharing game.
For both cases, the variance of price declines with less stringent power flow limits. 
Under a specific flow limit, the average nodal prices for the two cases are very close, while the variation of energy sharing price is smaller than that of LMP, which verifies effectiveness of energy sharing in reducing price discrimination.

\begin{figure}[h]
\centerline{\includegraphics[width=0.7\columnwidth]{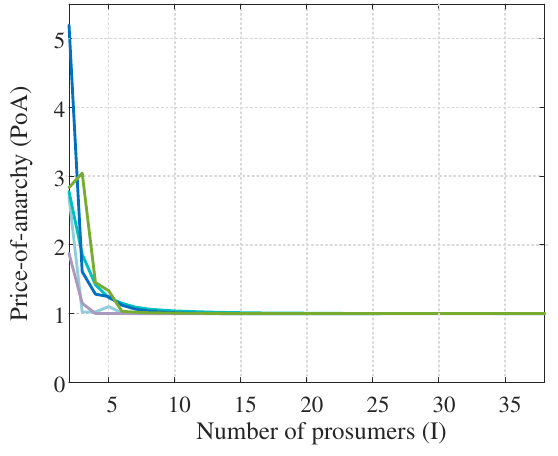}}
\setlength{\abovecaptionskip}{0pt}
\caption{Price of anarchy (PoA) across different prosumer number $I$.}
\label{fig:RWL}
\end{figure}

Next, we adjust the number $I$ of prosumers from $2$ to $38$. For each $I$, parameters $c_i, d_i$ are uniformly randomly chosen from $[0.001,0.01]$ and $[0.1,1.0]$, respectively, and 5 scenarios are sampled and tested. 
The price of anarchy (PoA) defined in \eqref{eq:RWLbounds} is shown in Fig. \ref{fig:RWL} across different $I$. 
We observe that in every scenario, PoA is no less than 1 and converges to 1 as $I$ grows, which verifies Proposition \ref{Thm:prop-5}.

\subsection{IEEE 69-bus test system}
\begin{figure}[h]
\centerline{\includegraphics[width=1.0\columnwidth]{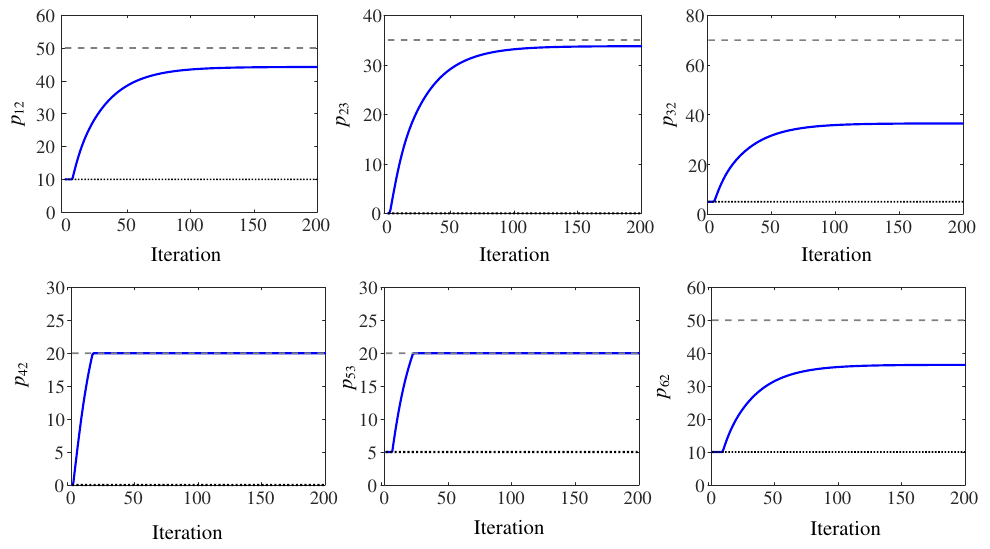}}
\setlength{\abovecaptionskip}{0pt}
\caption{Change of $p_i$ during iterations on the IEEE 69-bus system.}
\label{fig:IEEE69}
\end{figure}

To provide insights on more practical situations, we incorporate capacity limits into our bidding algorithm and test its performance on the IEEE 69-bus system. There are six prosumers with flexible net productions, located at nodes 12, 23, 32, 42, 53, 62, respectively. The prosumers at other nodes are inelastic, so the upper and lower bounds for $p_i$ are all set to zero. It takes around 200 iterations and 4.92s for the algorithm to converge. The change of $p_i$ of flexible prosumers during iterations are shown in Fig. \ref{fig:IEEE69}. The blue line refers to $p_i$, and the black and grey lines refer to the lower and upper capacity limits, respectively. This verifies the convergence of the bidding algorithm. Moreover, capacity limits are always satisfied. At the GNE, we have $p_{12}^*=44.28$kW, $p_{23}^*=33.71$kW, $p_{32}^*=36.52$kW, $p_{42}^*=20.00$kW, $p_{53}^*=20.00$kW, $p_{61}^*=36.49$kW, which are the same as the optimal solution of problem \eqref{eq:central} with the same parameters.

\begin{figure}[h]
\centerline{\includegraphics[width=0.7\columnwidth]{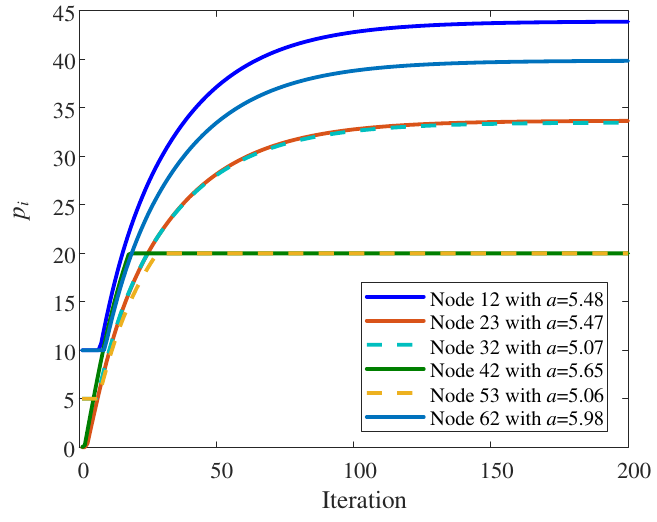}}
\setlength{\abovecaptionskip}{0pt}
\caption{Change of $p_i$ during iterations on the IEEE 69-bus system with heterogeneous $a$.}
\label{fig:hetera}
\end{figure}

Moreover, in the following, we replace the $a$ in the proposed model with heterogeneous $a_i$, randomly generate the $a_i$ for different nodes, and test the performance of the bidding algorithm. The change of $p_i$ during iterations are shown in Fig. \ref{fig:hetera}. We can find that the algorithm still converges. Moreover, we have checked that at the GNE, the $\bar p$ is the unique optimal solution of problem \eqref{eq:central}.

The computational time to reach an equilibrium of different systems are compared in Table \ref{tab:R1}. Since the bidding of prosumers can be run in parallel, the computational time does not change much with the growth of system size. In this paper, we focus on the real-time market. Since the proposed algorithm only takes a few seconds to converge, the frequency of market clearance can be up to every 5 min or even faster. In that small time resolution, the predictions of renewable generation and load demand can be very accurate.
\begin{table}[h]
	\renewcommand{\arraystretch}{1.3}
	\centering
	\caption{Computation time of different systems}
	\label{tab:R1}
	\begin{tabular}{cccc}
		\hline 
		& 2-prosumer system & IEEE 38-bus & IEEE 69-bus  \\
		\hline
		Time (s) & 1.13 & 1.58 & 4.92\\
		\hline
	\end{tabular}
\end{table}

\subsection{Comparison with previous work}
The proposed energy sharing mechanism is novel in that it incorporates price-making prosumers with endogenously given market roles, network constraints, and market power limitation. To the best of our knowledge, there is no existing method with all these three features that can be used for numerical comparison. Still, the advantages of the proposed method can be verified as follows:

1) \emph{Price-making prosumers with endogenously given market roles}. Compared with \cite{liu2017energy,Sharing-setprice2} that assume price-taking prosumers, the prosumers in this paper are price-makers. Moreover, the role of prosumer in our model is endogenously assigned based on the other prosumers’ situations, as in the example in TABLE \ref{tab:marketrole}. This is distinct from \cite{morstyn2018bilateral} which pre-divides participants into sellers and buyers.

2) \emph{Network constraints}. Despite some studies with price-making prosumers \cite{chen2020energy,chen2020approaching}, network constraints are seldom considered due to the theoretical complexity. This paper proposes a novel market clearing rule \eqref{eq:new-rule} to ensure fairness of the market while adhering to network constraints. The proposed model is practical with network constraints known only to the platform and individual constraint only to each prosumer.

3) \emph{Market Power limitation}. Potential market power exploitation is discussed in Section \ref{sec-3.2}, and we propose a price regulation policy \eqref{eq:priceregulate} to mitigate the market power. This has not been reported in previous work.

In particular, the proposed mechanism differs from that in \cite{johari2011parameterized} in that: 1) The market in \cite{johari2011parameterized} is designed only for sellers while our market works for both sellers and buyers. 2) The proposed market has a generalized Nash equilibrium when $I \ge 1$, while the Nash equilibrium in \cite{johari2011parameterized} exists only when $N \ge 2$. Here, both $I$ and $N$ represent the number of market participants. 3) Network constraints and market power mitigation were not considered in \cite{johari2011parameterized}.

\section{Conclusion}
\label{sec-7}
The proliferation of prosumers with distributed generation facilities calls for new business models for energy management. Energy sharing is one of those business models that has great potential. A well-designed energy sharing mechanism is imperative. This paper comes up with an energy sharing mechanism considering network constraints and fairness of prices. Price regulation is introduced to limit market power, ensuring the existence of a unique and socially near-optimal market equilibrium. Several advantageous properties of the improved energy sharing market equilibrium are disclosed. A practical bidding algorithm with its convergence condition is developed to reach the designed market equilibrium. As revealed in this paper, traditional electricity markets and sharing markets have numerous features in common. Characterization of their similarity and difference shall deepen our understanding of energy sharing, which will be our future research direction.

\ifCLASSOPTIONcaptionsoff
\newpage
\fi

\bibliographystyle{IEEEtran}
\bibliography{IEEEabrv,mybib}

\appendices

\makeatletter
\@addtoreset{equation}{section}
\@addtoreset{theorem}{section}
\makeatother
\setcounter{definition}{0} 
\renewcommand{\theequation}{A.\arabic{equation}}
\renewcommand{\thetheorem}{A.\arabic{theorem}}
\renewcommand{\thedefinition}{A\arabic{definition}}

\section{Proof of Proposition \ref{Thm:prop-0}}
\label{apen-0}

First, we prove that a unique optimal solution of \eqref{eq:AGG} exists. The construction of the $\pi_{il}$ matrix after equation \eqref{eq:flowlimits} implies boundedness of the feasible set \eqref{eq:AGG.2}--\eqref{eq:AGG.2} for $p$. Indeed, that construction leads to:
    \begin{eqnarray}
    - \tilde F \leq -B\tilde C^T(\tilde C B \tilde C^T)^{-1} (D_{-I}- p_{-I})\leq F\nonumber
    \end{eqnarray}
    where $\tilde F:=\{\tilde F_l,~\forall l\in\mathcal{L} \}$, $F:=\{F_l,~\forall l\in\mathcal{L} \}$. $D_{-I}:=\{D_1,...,D_{I-1}\}$and $ p_{-I}:=\{p_1,...,p_{I-1}\}$ are obtained by removing the last prosumer from vectors $D$ and $p$. Multiplying both sides by $\tilde C$ implies boundedness of $p_{-I}$, and further boundedness of $p$ due to $\sum_{i=1}^I p_i = \sum_{i=1}^{I} D_i $ being a constant. 
    Hence, the strictly convex objective function \eqref{eq:AGG.1} attains a unique optimal solution in the compact convex set \eqref{eq:AGG.2}--\eqref{eq:AGG.3}.

Problem \eqref{eq:AGG} can be rewritten as
\bsq
\label{eq:AGG-eq}
\begin{align}
    \mathop{\min}_{p_i,  q_i, \forall i \in \mathcal{I}} ~& \sum \limits_{i=1}^I (c_ip_i^2+d_ip_i) \label{eq:AGG-eq.1}\\
    \mbox{s.t.}~ & \sum \limits_{i=1}^I q_i=0 \label{eq:AGG-eq.2}\\
    ~ & q_i=D_i-p_i:\lambda_i, \forall i \in \mathcal{I} \label{eq:AGG-eq.3}\\
    ~ & -\tilde F_l \le \sum \limits_{i=1}^I \pi_{il}q_i \le F_l,\forall l \in \mathcal{L} \label{eq:AGG-eq.4}
\end{align}
\esq
Denote $\mathcal{Q}$ as the feasible region of $q=(q_i,~i \in \mathcal{I})$ characterized by \eqref{eq:AGG-eq.2} and \eqref{eq:AGG-eq.4}. 
$\mathcal{Q}$ is a closed convex set. The Lagrangian function of problem \eqref{eq:AGG-eq} is
\begin{align}
    \mathbb{L}(p,q,\lambda)=\sum \limits_{i=1}^I (c_ip_i^2+d_ip_i)-\sum \limits_{i=1}^I \lambda_i(q_i+p_i-D_i)
\end{align}
defined on $\Omega:=\mathbb{R}^I \times \mathcal{Q} \times \mathbb{R}^I$.

Let $(\tilde{p},\tilde{q},\tilde{\lambda})$ be a saddle point of $\mathbb{L}(p,q,\lambda)$, then $(\tilde{p},\tilde{q},\tilde{\lambda}) \in \Omega$, and every $(p,q,\lambda) \in \Omega$ satisfies:
\begin{align}
\label{eq:AGG-condition1}
\left\{
\begin{array}{l}
     J_i(p_i)-J_i(\tilde{p}_i)-(p_i-\tilde{p}_i)\tilde{\lambda}_i \ge 0,\forall i \in \mathcal{I}  \\
     -\sum \limits_{i=1}^I (q_i-\tilde{q}_i)\tilde{\lambda}_i \ge 0 \\
     \sum \limits_{i=1}^I (\lambda_i-\tilde{\lambda}_i)(\tilde{q}_i+\tilde{p}_i-D_i) \ge 0
\end{array}
\right.
\end{align}

Suppose $(\bar p,\bar b, \bar \lambda)$ is a CE of the sharing game, then $\bar \lambda_i=\frac{1}{a}(\bar b_i+\bar p_i-D_i)$, $\bar q_i = D_i-\bar p_i$, $\forall i \in \mathcal{I}$. 
By the optimality of $(\bar p_i, \bar b_i)$ for \eqref{eq:competitiveequilibrium}, we have for all $p \in \mathbb{R}^I$:
\begin{align}
\label{eq:AGG-condition2}
    J_i(p_i)-J_i(\bar p_i)-(p_i-\bar p_i)\bar \lambda_i \ge 0,\forall i \in \mathcal{I}
\end{align}

Problem \eqref{eq:new-rule} is equivalent to
\bsq
\label{eq:market-rule2}
\begin{align}
    \mathop{\min}_{q_i,\forall i \in \mathcal{I}}~ & \sum \limits_{i=1}^I (q_i-b_i)^2 \\
    \mbox{s.t.}~ & \sum \limits_{i=1}^I q_i = 0 \\
    ~ & -\tilde F_l \le \sum \limits_{i=1}^I \pi_{il}q_i \le F_l, ~\forall l \in \mathcal{L}
\end{align}
\esq
Given $\bar b$, the optimal solution of \eqref{eq:market-rule2} is $\bar q$, which satisfies the first-order necessary condition for optimality:
\begin{align}
\label{eq:AGG-condition3}
    \sum \limits_{i=1}^I (q_i-\bar q_i)(\bar q_i-\bar b_i) \ge 0,\forall q \in \mathcal{Q}
\end{align}

If $(\bar p,\bar b, \bar \lambda)$ is a CE, we have $\bar b_i=D_i-\bar p_i+a \bar \lambda_i$ for all $i \in \mathcal{I}$. Let $\tilde{p}_i = \bar p_i$, $\tilde{\lambda}_i=\bar \lambda_i$, $\tilde{q}_i = -a\bar \lambda_i +\bar b_i$, $\forall i \in \mathcal{I}$. Then $(\tilde{p},\tilde{q},\tilde{\lambda})$ satisfies \eqref{eq:AGG-condition1} and hence $\tilde{p}$ is the unique optimal solution of \eqref{eq:AGG} or \eqref{eq:AGG-eq}. Therefore, $\bar q_i=D_i-\bar p_i = D_i-\tilde p_i$ is unique for all $i \in \mathcal{I}$. The dual optimal $\bar \lambda$ of \eqref{eq:AGG-eq} is also unique, which implies uniqueness of $\bar b_i = a \bar \lambda_i + \bar q_i$  for all $i \in \mathcal{I}$.

In the other direction, suppose $(\tilde{p},\tilde{q},\tilde{\lambda})$ is the unique primal-dual optimal solution of \eqref{eq:AGG-eq}. Let $\bar p_i=\tilde{p}_i$, $\bar \lambda_i = \tilde{\lambda}_i$, $\bar b_i =D_i-\tilde{p}_i+a\tilde{\lambda}_i$ for all $i \in \mathcal{I}$, and then we can check that $(\bar p,\bar b, \bar \lambda)$ satisfies \eqref{eq:AGG-condition2} and \eqref{eq:AGG-condition3} and is thus a CE.

\setcounter{definition}{0} 
\renewcommand{\theequation}{B.\arabic{equation}}
\renewcommand{\thetheorem}{B.\arabic{theorem}}
\renewcommand{\thedefinition}{B\arabic{definition}}
\section{Proof of Proposition \ref{Thm:prop-1}}
\label{apen-1}
\emph{Part 1: Characterization of GNE and VE.}
A1 implies feasibility of convex problem \eqref{eq:new-rule} where all the constraints are affine and thus Slater's condition is satisfied. Therefore, \eqref{eq:new-rule} attains a dual optimal point with zero duality gap, and the following KKT condition is necessary and sufficient for $(\lambda; \eta, \alpha^\pm)$ to be a primal-dual optimal point of \eqref{eq:new-rule}:
\bsq
\label{eq:lowerKKT}
\begin{align}
 2\lambda_i+a\eta+a\sum \limits_{l=1}^L \pi_{il}\alpha_l^{-}-a\sum \limits_{l=1}^L \pi_{il}\alpha_l^{+}=0, \forall i \in \mathcal{I} \label{eq:lowerKKT.1}\\ 
  \sum \limits_{i=1}^I (a\lambda_i-b_i)=0 \label{eq:lowerKKT.2}\\
  0 \le \left(\sum \limits_{i=1}^I \pi_{il} (-a\lambda_i+b_i)+\tilde F_l\right) \perp \alpha_l^{-} \ge 0,\forall l \in \mathcal{L} \label{eq:lowerKKT.3}\\
  0 \le \left(-\sum \limits_{i=1}^I \pi_{il} (-a\lambda_i+b_i)+F_l\right) \perp \alpha_l^{+} \ge 0,\forall l \in \mathcal{L} \label{eq:lowerKKT.4}
\end{align}
\esq

By strict convexity of \eqref{eq:new-rule.1}, the primal optimal $\lambda$ satisfying \eqref{eq:lowerKKT} is unique. Moreover, if the energy sharing network is connected and radial, then by Lemma \ref{lemma:LICQ} below, we know that the dual optimal $(\eta,\alpha^{\pm})$ satisfying \eqref{eq:lowerKKT} is also unique. 
\begin{lemma}
\label{lemma:LICQ}
If the energy sharing network is connected and radial, then problem \eqref{eq:new-rule} satisfies the linear independence constraint qualification (LICQ).
\end{lemma} 

\begin{proof}

If the network is radial, then $L=I-1$, and it can be verified that all the $L$ columns of $\Pi$, together with $\mathbf{1}_{I}$, are linearly independent. Therefore, the gradient vectors of equality and active inequality constraints in \eqref{eq:new-rule} at any feasible point are linearly independent, i.e., LICQ is satisfied.
\end{proof}




Given that problem \eqref{eq:new-rule} satisfies LICQ, problem \eqref{eq:sharing-game} for each prosumer $i\in \mathcal{I}$ can be equivalently reformulated as:
\begin{align}
    \mathop{\min}_{b_i,\lambda, \eta, \alpha^{\pm}} ~ & c_i(D_i\!+\!a\lambda_i\!-\!b_i)^2+d_i(D_i\!+\!a\lambda_i\!-\!b_i) + \lambda_i(-a\lambda_i\!+\!b_i)  \nonumber\\
    \mbox{s.t.}~ &  \eqref{eq:lowerKKT.1}-\eqref{eq:lowerKKT.4} \label{eq:sharing-eq}
\end{align}
which is a mathematical problem with equilibrium constraint (MPEC). It has been proved that an MPEC violates most of the common constraint qualifications, such as LICQ, at any feasible point \cite{su2004equilibrium}, and therefore the KKT condition is generally not applicable to characterize its optimal point.

For every prosumer $i\in \mathcal{I}$, given any feasible point $(\bar b_i, \bar \lambda, \bar \eta, \bar \alpha^{\pm})$ of MPEC \eqref{eq:sharing-eq} and any $\bar b_{-i}$, we define the following index sets of active and inactive constraints:
\begin{align}
\mathbb{I}_1:=~ &\{l \in \mathcal{L}~|~\tilde F_l+\sum \limits_{i=1}^I \pi_{il}(-a\bar \lambda_i+\overline b_i)=0<\bar \alpha_l^{-}\} \nonumber\\
\mathbb{J}_1:=~ &\{l\in \mathcal{L}~|~\tilde F_l+\sum \limits_{i=1}^I \pi_{il}(-a\bar \lambda_i+\overline b_i)=0=\bar \alpha_l^{-}\} \nonumber\\
\mathbb{K}_1:=~ &\{l\in \mathcal{L}~|~\tilde F_l+\sum \limits_{i=1}^I \pi_{il}(-a\bar \lambda_i+\overline b_i)>0= \bar \alpha_l^{-}\} \nonumber\\
\mathbb{I}_2:= ~ & \{l\in \mathcal{L}~|~F_l-\sum \limits_{i=1}^I \pi_{il}(-a\bar \lambda_i+\overline b_i)=0 < \bar \alpha_l^{+}\} \nonumber\\
\mathbb{J}_2:=~ &\{l\in \mathcal{L}~|~F_l-\sum \limits_{i=1}^I \pi_{il}(-a\bar \lambda_i+\overline b_i)=0=\bar \alpha_l^{+}\} \nonumber\\
\mathbb{K}_2:=~ &\{l\in \mathcal{L}~|~F_l-\sum \limits_{i=1}^I \pi_{il}(-a\bar \lambda_i+\overline b_i)>0= \bar \alpha_l^{+}\} \nonumber
\end{align}

Associated with the specific point $(\bar b_i, \bar \lambda, \bar \eta, \bar \alpha^{\pm})$ above, we construct a relaxed nonlinear program (RNLP):
\bsq
\label{eq:relax-NP}
\begin{align}
    \mathop{\min}_{b_i,\lambda, \eta, \alpha^{\pm}} ~ & c_i(D_i\!+\!a\lambda_i\!-\!b_i)^2+d_i(D_i\!+\!a\lambda_i\!-\!b_i) + \lambda_i(-a\lambda_i\!+\!b_i) \nonumber\\
    \mbox{s.t.}~ &  \eqref{eq:lowerKKT.1}-\eqref{eq:lowerKKT.2} \label{eq:relax-NP.2}\\
    ~ &  \tilde F_l+\sum \limits_{i=1}^I \pi_{il}(-a\lambda_i+b_i)=0, \alpha_l^{-} \ge 0, \forall l \in \mathbb{I}_1 \label{eq:relax-NP.3}\\
    ~ &  \tilde F_l+\sum \limits_{i=1}^I \pi_{il}(-a\lambda_i+b_i) \ge 0, \alpha_l^{-} \ge 0, \forall l \in \mathbb{J}_1  \label{eq:relax-NP.4}\\
    ~ &  \tilde F_l+\sum \limits_{i=1}^I \pi_{il}(-a\lambda_i+b_i) \ge 0, \alpha_l^{-}= 0, \forall l \in \mathbb{K}_1 \label{eq:relax-NP.5}\\
    ~ &  F_l-\sum \limits_{i=1}^I \pi_{il}(-a\lambda_i+b_i)=0, \alpha_l^{+} \ge 0, \forall l \in \mathbb{I}_2 \label{eq:relax-NP.6}\\
    ~ &  F_l-\sum \limits_{i=1}^I \pi_{il}(-a\lambda_i+b_i) \ge 0, \alpha_l^{+} \ge 0, \forall l \in \mathbb{J}_2  \label{eq:relax-NP.7}\\
    ~ &  F_l-\sum \limits_{i=1}^I \pi_{il}(-a\lambda_i+b_i) \ge 0, \alpha_l^{+}= 0, \forall l \in \mathbb{K}_2  \label{eq:relax-NP.8}
\end{align}
\esq


\begin{lemma}
Suppose the energy sharing network is radial, and $\mathbb{J}_1 \!\cup\! \mathbb{J}_2 \!= \!\emptyset$ at $(\bar b_i, \!\bar \lambda,\! \bar \eta, \!\bar \alpha^{\pm})$.\footnote{We skipped the condition $\mathbb{J}_1 \cup \mathbb{J}_2 = \emptyset$ in Proposition \ref{Thm:prop-1} for conciseness and also because it holds almost surely in practice.} Then RNLP \eqref{eq:relax-NP} satisfies LICQ at $(\bar b_i, \bar \lambda, \bar \eta, \bar \alpha^{\pm})$. 
\end{lemma}

\begin{proof}
Recall $\Pi$ is the matrix of line flow distribution factors constructed in the preceding texts. For a particular prosumer $i \in \mathcal{I}$, consider its RNLP \eqref{eq:relax-NP}. At the point $(\bar b_i, \bar \lambda, \bar \eta, \bar \alpha^{\pm})$ under consideration, the line index set $\mathcal{L}$ is composed of five mutually exclusive sets: $\mathbb{I}_1$, $\mathbb{J}_1$, $\mathbb{I}_2$, $\mathbb{J}_2$, and $\mathbb{K}_1 \cap \mathbb{K}_2$. 
 
\begin{itemize}
\item The gradient vector of equality constraint \eqref{eq:lowerKKT.2} is:
\begin{eqnarray}\nonumber
\vec{e}_0 = \left[-1; ~a,\dots,a;~ 0;~ 0,\dots,0;~ 0,\dots,0\right]^T 
\end{eqnarray}
with components corresponding to $(b_i; \lambda; \eta; \alpha^-; \alpha^+)$. 

\item The gradient vectors of \eqref{eq:lowerKKT.1} for $j \in \mathcal{I}$ are:
\begin{eqnarray} 
\vec{e}_{1,j} &=& [0; ~0,...,2,...,0;~ a;~ a\pi_{j1},...,a\pi_{jL}; \nonumber \\
&& \quad ~-a\pi_{j1},...,-a\pi_{jL} ]^T  \nonumber
\end{eqnarray}
where the term ``$2$'' appears only at the $j$-th location of the subvector corresponding to $\lambda$. 

\item For $l \in \mathbb{I}_1 \cup \mathbb{J}_1 \cup \mathbb{I}_2 \cup \mathbb{J}_2 =: \mathcal{L}_2$, flow constraints $F_l \pm \sum_{j=1}^I \pi_{jl}(-a  \lambda_j+ b_j) \geq 0~ (\textnormal{or}=0)$ that are equality and \emph{active} inequality at $(\bar \lambda, \bar b_i)$ have gradient vectors:
\begin{eqnarray}\nonumber
\vec{e}_{2,l} = \left[\pi_{il}; ~-a\pi_{1l},\dots,-a \pi_{Il};~ 0;~ 0,\dots,0;~ 0,\dots,0\right]^T 
\end{eqnarray}

\item For $l \in (\mathbb{K}_1 \cap \mathbb{K}_2) \cup \mathbb{J}_1  \cup \mathbb{J}_2 \cup \mathbb{I}_2 = \mathbb{J}_1 \cup \mathbb{K}_1=:\mathcal{L}_{3-}$, constraints $\alpha_l^- \geq 0~ (\textnormal{or}=0)$ that are equality and \emph{active} inequality at $\bar \alpha^-$ have gradient vectors:
\begin{eqnarray}\nonumber
\vec{e}_{3-,l} = \left[0; ~0,\dots,0;~ 0;~ 0,...,1,...,0;~ 0,\dots,0\right]^T 
\end{eqnarray}
where the term ``$1$'' appears only at the $l$-th location of the subvector corresponding to $\alpha^-$. 

\item For $l \in (\mathbb{K}_1 \cap \mathbb{K}_2) \cup \mathbb{J}_1  \cup \mathbb{J}_2 \cup \mathbb{I}_1 = \mathbb{J}_2 \cup \mathbb{K}_2=:\mathcal{L}_{3+}$, constraints $\alpha_l^+ \geq 0~ (\textnormal{or}=0)$ that are equality and \emph{active} inequality at $\bar \alpha^+$ have gradient vectors:
\begin{eqnarray}\nonumber
\vec{e}_{3+,l} = \left[0; ~0,\dots,0;~ 0;~ 0,\dots,0;~ 0,...,1,...,0\right]^T 
\end{eqnarray}
where the term ``$1$'' appears only at the $l$-th location of the subvector corresponding to $\alpha^+$. 
\end{itemize}

We next find coefficients $k_0$, $k_1:=(k_{1,j},~\forall j\in \mathcal{I})$, $k_2:=(k_{2,l},~\forall l\in\mathcal{L}_2)$, $k_{3-}:=(k_{3-,l},~\forall l\in\mathcal{L}_{3-})$, $k_{3+}:=(k_{3+,l},~\forall l\in\mathcal{L}_{3+})$, where $k_1$, $k_2$, $k_{3-}$, $k_{3+}$ are column vectors, such that
\begin{eqnarray} \nonumber
&& \quad k_0 \vec{e}_0 +\sum_{j\in\mathcal{I}} k_{1,j} \vec{e}_{1,j} + \sum_{l\in\mathcal{L}_2} k_{2,l} \vec{e}_{2,l} \\
&& + \sum_{l\in\mathcal{L}_{3-}} k_{3-,l} \vec{e}_{3-,l} + \sum_{l\in\mathcal{L}_{3+}} k_{3+,l} \vec{e}_{3+,l}=0 \label{eq:RNLP-linear-dependence}
\end{eqnarray}  

\begin{itemize}
\item Elements in \eqref{eq:RNLP-linear-dependence} corresponding to $b_i$ satisfy:
\begin{eqnarray}
-k_0 + \pi_i^{\mathcal{L}_2} k_2 = 0 \label{eq:RNLP-linear-dependence:bi}
\end{eqnarray} 
where $\pi_i$ is the $i$-th row of line flow distribution factor matrix $\Pi$ and superscript $\mathcal{L}_2$ means taking the submatrix (subvector) by retaining only the columns in $\mathcal{L}_2$.

\item Elements in \eqref{eq:RNLP-linear-dependence} corresponding to $\lambda$ satisfy:
\begin{eqnarray}
a k_0 \mathbf{1}_{I} + 2 k_1 -a \Pi^{\mathcal{L}_2} k_2 = 0 \label{eq:RNLP-linear-dependence:lambda}
\end{eqnarray} 
where $\mathbf{1}_{I}$ is the $I$-dimensional column vector of all ones.  

\item Elements in \eqref{eq:RNLP-linear-dependence} corresponding to $\eta$ satisfy:
\begin{eqnarray}
a \mathbf{1}_{I}^T k_1 = 0 \label{eq:RNLP-linear-dependence:eta}
\end{eqnarray} 
 
\item Elements in \eqref{eq:RNLP-linear-dependence} corresponding to $\alpha^{\pm}$ satisfy:
\begin{eqnarray}
a \Pi^T k_1 &=& - [\dots, k_{3-,l}, \dots,0,\dots]^T  \nonumber \\
a \Pi^T k_1 & = & [\dots, k_{3+,l}, \dots,0,\dots]^T \nonumber
\end{eqnarray}  
where in the first line the $l$-th element of the right-hand-side vector is $k_{3-,l}$ if $l\in \mathcal{L}_{3-}$ and zero otherwise; in the second line the $l$-th element of the right-hand-side vector is $k_{3+,l}$ if $l\in \mathcal{L}_{3+}$ and zero otherwise. We imply:
\begin{eqnarray}
 k_{3-,l}&=& 0,\quad \forall  l \in \mathbb{I}_2 \nonumber\\
 k_{3+,l}&=&0, \quad \forall l \in \mathbb{I}_1 \nonumber \\
-k_{3-,l} &=& k_{3+,l}=:k_{3,l},~ \forall l \in (\mathbb{K}_1 \!\cap\! \mathbb{K}_2) \!\cup \!\mathbb{J}_1  \!\cup \!\mathbb{J}_2  =:\mathcal{L}_3 \nonumber
\end{eqnarray}
Let $k_3$ denote the $|\mathcal{L}_3|$-dimensional column vector $(k_{3,l},~\forall l\in\mathcal{L}_3)$, and $\hat k_3$ the $L$-dimensional column vector whose $l$-th element is $k_{3,l}$ if $l\in\mathcal{L}_3$ and zero otherwise. Then we have
\begin{eqnarray}
a \Pi^T k_1 = \hat k_3 \label{eq:RNLP-linear-dependence:alpha}
\end{eqnarray} 
\end{itemize}

By \eqref{eq:RNLP-linear-dependence:bi}, \eqref{eq:RNLP-linear-dependence:lambda}, we have 
\begin{eqnarray}
2k_1 = a\left(\Pi^{\mathcal{L}_2} - \Pi_i^{\mathcal{L}_2} \right) k_2\label{eq:proof-LICQ-1}
\end{eqnarray} 
where $\Pi_i$ is the $I\times L$ matrix whose every row is $\pi_i$. By \eqref{eq:RNLP-linear-dependence:eta}, \eqref{eq:RNLP-linear-dependence:alpha}, we have:
\begin{eqnarray}
a (\Pi - \Pi_i)^T k_1 = \hat k_3 \label{eq:proof-LICQ-2}
\end{eqnarray} 
Note that the matrix $(\Pi - \Pi_i)$ has its $i$-th row zero and other $(I-1)$ rows linearly independent. Hence, we remove the $i$-th row of $(\Pi - \Pi_i)$ and denote the remaining $(I-1)\times L$ matrix as $\tilde \Pi$ (which has a different meaning from $\tilde \Pi$ in the preceding texts). Correspondingly, we remove $k_{1,i}$ and denote the remaining $(I-1)$-dimensional column vector as $\tilde k_1$.  

By \eqref{eq:proof-LICQ-1}, we further have:
\begin{eqnarray}
2\tilde k_1 &=& a \tilde \Pi^{\mathcal{L}_2} k_2 \label{eq:proof-LICQ-3} \\
k_{1,i} &=& 0 \label{eq:proof-LICQ-4}
\end{eqnarray}
By \eqref{eq:proof-LICQ-2}, we have:
\begin{eqnarray}
a\tilde \Pi^T \tilde k_1 &=& \hat k_3 \label{eq:proof-LICQ-5}  
\end{eqnarray} 
By \eqref{eq:RNLP-linear-dependence:eta}, \eqref{eq:proof-LICQ-4}, we have:
\begin{eqnarray}
\mathbf{1}_{I-1}^T \tilde k_1 &=&  0 \label{eq:proof-LICQ-6}  
\end{eqnarray} 

If the network is radial, then $L=I-1$, and $\tilde \Pi$ is $(I-1)\times (I-1)$ invertible square matrix. In this case, combining \eqref{eq:proof-LICQ-3} and \eqref{eq:proof-LICQ-5}, we have:
\begin{eqnarray}
\tilde k_1 &=& \frac{1}{a} \tilde \Pi^{-T} \hat k_3 \nonumber \\ 
 &=& \frac{1}{a} \left(\tilde \Pi^{-T}\right)^{\mathcal{L}_3}  k_3 = \frac{a}{2} \tilde \Pi^{\mathcal{L}_2} k_2 \label{eq:proof-LICQ-final}  
\end{eqnarray}
where $\left(\tilde \Pi^{-T}\right)^{\mathcal{L}_3}$ is the submatrix of $\tilde \Pi^{-T}$ by retaining only the columns in $\mathcal{L}_3$. 

If $\mathbb{J}_1 \cup \mathbb{J}_2 = \emptyset$ at $(\bar b_i, \bar \lambda, \bar \eta, \bar \alpha^{\pm})$, then $\mathcal{L}_2 = \mathbb{I}_1 \cup \mathbb{I}_2 $ and $\mathcal{L}_3 = \mathbb{K}_1 \cap \mathbb{K}_2$ are mutually exclusive. Therefore, the column spaces of $\left(\tilde \Pi^{-T}\right)^{\mathcal{L}_3}$ and $\Pi^{\mathcal{L}_2}$ are orthogonal, so that \eqref{eq:proof-LICQ-final} implies $k_2=0$ and $k_3=0$. It is then straightforward to imply that all the coefficients in \eqref{eq:RNLP-linear-dependence} are zero and hence LICQ holds.
\end{proof}


By Theorem 5.18 in \cite{fukushima1980theory}, if RNLP \eqref{eq:relax-NP} satisfies LICQ and $(\bar b_i, \bar \lambda, \bar \eta, \bar \alpha^{\pm})$ is a local optimal point of \eqref{eq:sharing-eq}, then there exists a \emph{unique} dual optimal $(\bar \epsilon,\bar \delta,\bar \gamma^{\pm},\bar \phi^{\pm})$ that satisfies:
\bsq
\label{eq:KKT-MPEC}
\begin{align}
    -2c_i (D_i+a\bar \lambda_i-\bar b_i)-d_i+\bar \lambda_i-\bar \delta \quad &~ \nonumber\\
    -\sum \limits_{l=1}^L \pi_{il}\bar \gamma_l^{-}+\sum \limits_{l=1}^L \pi_{il} \bar \gamma_l^{+}  = 0 & \label{eq:KKT-MPEC.1}\\
    2ac_i(D_i+a\bar \lambda_i-\bar b_i)+ad_i-2a\bar \lambda_i+\bar b_i+2\bar \epsilon_i \quad &~ \nonumber\\
    + a\bar \delta + a\sum \limits_{l=1}^L \pi_{il} \bar \gamma_l^{-} - a \sum \limits_{l=1}^L \pi_{il} \bar \gamma_l^{+} =0 & \label{eq:KKT-MPEC.2}\\
    \forall j \!\in\! \mathcal{I}\backslash\{i\}: 2\bar \epsilon_j \!+\!a\bar \delta \!+\! a\sum \limits_{l=1}^L \pi_{jl} \bar\gamma_l^{-}\! -\! a \sum \limits_{l=1}^L \pi_{jl} \bar \gamma_l^{+}  =0 & \label{eq:KKT-MPEC.3}\\
    a\sum \limits_{i=1}^I \bar\epsilon_i  =0  & \label{eq:KKT-MPEC.4}\\
     \forall l \in \mathcal{L}: a  \sum \limits_{i=1}^I \pi_{il}\bar \epsilon_i-\bar\phi_l^{-}  =0 & \label{eq:KKT-MPEC.5}\\
    \forall l \in \mathcal{L}: - a \sum \limits_{i=1}^I \pi_{il}\bar \epsilon_i-\bar\phi_l^{+}  =0  & \label{eq:KKT-MPEC.6}\\
\forall l\in\mathbb{J}_1 \!\cup\! \mathbb{K}_1: 0 \!\le \! \left(\tilde F_l \!+\!\sum \limits_{i=1}^I \pi_{il} (-a\bar\lambda_i\!+\!\bar b_i)\right) \!\perp\! \bar\gamma_l^{-} \!\ge\! 0& \label{eq:KKT-MPEC.complementary-gamma-minus}\\
\forall l\in\mathbb{I}_1 \cup \mathbb{J}_1: ~ 0 \!\le \bar\alpha_l^- \perp \bar\phi_l^{-} \ge\! 0 & \label{eq:KKT-MPEC.complementary-phi-minus}\\
\forall l\in\mathbb{J}_2 \!\cup\! \mathbb{K}_2: 0 \!\le\! \left(F_l \! - \!\sum \limits_{i=1}^I \pi_{il} (-a\bar\lambda_i\!+\!\bar b_i)\right)\! \perp\! \bar\gamma_l^{+} \!\ge \!0 & \label{eq:KKT-MPEC.complementary-gamma-plus}\\
\forall l\in\mathbb{I}_2 \cup \mathbb{J}_2: ~ 0 \!\le \bar\alpha_l^+ \perp \bar\phi_l^{+} \ge\! 0 & \label{eq:KKT-MPEC.complementary-phi-plus}\\
\forall l\in\mathbb{I}_1: ~\tilde F_l + \sum \limits_{i=1}^I \pi_{il} (-a\bar\lambda_i +\bar b_i ) = 0 & \label{eq:KKT-MPEC.primal-feasibility-flowminus}\\
\forall l\in\mathbb{I}_2: ~F_l - \sum \limits_{i=1}^I \pi_{il} (-a\bar\lambda_i +\bar b_i ) = 0 & \label{eq:KKT-MPEC.primal-feasibility-flowplus}\\
 \forall l\in\mathbb{K}_1: ~\bar\alpha_l^- = 0; \quad  \forall l\in\mathbb{K}_2: ~\bar\alpha_l^+ = 0 & \label{eq:KKT-MPEC.primal-feasibility-alpha}\\
 \forall i \in \mathcal{I}: 2\bar \lambda_i+a\bar \eta+a\sum \limits_{l=1}^L \pi_{il}\bar \alpha_l^{-}-a\sum \limits_{l=1}^L \pi_{il}\bar \alpha_l^{+}=0 & \label{eq:KKT-MPEC.14}\\ 
  \sum \limits_{i=1}^I (a\bar \lambda_i-\bar b_i)  =0 &   \label{eq:KKT-MPEC.15} 
\end{align}
\esq

A GNE $(\bar p,\bar d)$ of the energy sharing game $\mathcal{G}$ is a solution of an equilibrium problem with equilibrium constraint (EPEC) \cite{su2004equilibrium} composed of $I$ groups of conditions \eqref{eq:KKT-MPEC} across $i \in \mathcal{I}$, which characterize local optimal points of $I$ different MPECs \eqref{eq:sharing-eq}, if the RNLP \eqref{eq:relax-NP} corresponding to each MPEC satisfies LICQ. At every GNE, we have:
\bsq
\begin{align}
    (\bar b_i, \bar \lambda, \bar \eta, \bar \alpha^{\pm}) &~ \in \mbox{\textbf{SOL}}(\mbox{MPEC}(\bar b_{-i})) ,\forall i \in \mathcal{I}\\
    \bar p_i &~ = D_i+a\bar \lambda_i - \bar b_i,\forall i \in \mathcal{I}
\end{align}
\esq
where \textbf{SOL}(.) denotes the optimal solution set.  

 Across the $I$ different MPECs \eqref{eq:sharing-eq} and their respective RNLPs \eqref{eq:relax-NP}, individual dual vectors satisfying \eqref{eq:KKT-MPEC} may or may not be identical, although they are all denoted by $(\bar \epsilon,\bar \delta, \bar \gamma^{\pm}, \bar \phi^{\pm})$ for conciseness. We make the following definition for the case in which these dual vectors are identical. 

\begin{definition} (Variational Equilibrium)
\label{def:VE}
A GNE $(\bar p,\bar d)$ of the energy sharing game $\mathcal{G}$ is a \emph{variational equilibrium} (VE) if at this equilibrium, there is an identical dual vector $(\bar \epsilon, \bar \delta, \bar \gamma^{\pm}, \bar \phi^{\pm})$ that satisfies \eqref{eq:KKT-MPEC} across all the prosumers $i \in \mathcal{I}$.
\end{definition}

\emph{Part 2: Proof for uniqueness of VE.} If A1 holds, problem \eqref{eq:central} has a unique optimal solution $\bar p$ due to strict convexity of objective \eqref{eq:central.1}.  Moreover, \eqref{eq:central} satisfies Slater's condition and attains a dual optimal point $(\bar \kappa, \bar \tau^{\pm})$ with zero duality gap, and the following KKT condition holds for $(\bar p; \bar \kappa, \bar \tau^{\pm})$:
\bsq
\label{eq:central-KKT}
\begin{align}
 2c_i \bar p_i +d_i -\frac{D_i -\bar p_i}{a(I-1)}+\bar \kappa \qquad \qquad  & \nonumber 
\\ 
 +\sum \limits_{l=1}^L \pi_{il}(\bar \tau_l^{-}-\bar \tau_l^{+})=0,~\forall i \in \mathcal{I} & \label{eq:central-KKT.1}\\
\sum \limits_{i=1}^I \bar p_i =\sum \limits_{i=1}^I D_i\\
0 \le \left(\sum \limits_{i=1}^I \pi_{il}(D_i-\bar p_i)+\tilde F_l\right) \perp \bar \tau_l^{-} \ge 0,~\forall l\in \mathcal{L} & \\
0 \le \left(-\sum \limits_{i=1}^I \pi_{il}(D_i-\bar p_i)+F_l\right) \perp \bar \tau_l^{+} \ge 0,~\forall l\in\mathcal{L} &
\end{align}
\esq

By Definition \ref{def:VE}, if $(\bar p,\bar b)$ is a VE, then there exist a vector $(\bar \lambda, \bar b, \bar \eta, \bar \alpha^{\pm})$ and a vector $(\bar \epsilon, \bar \delta, \bar \gamma^{\pm}, \bar \phi^{\pm})$ that together satisfy \eqref{eq:KKT-MPEC} and $\bar p_i = D_i+ a \bar \lambda_i-\bar b_i$ for all  $i\in\mathcal{I}$.
For every particular $i \in \mathcal{I}$, we add \eqref{eq:KKT-MPEC.2} with \eqref{eq:KKT-MPEC.3} for all $j \in \mathcal{I} \backslash\{i\}$ and combine \eqref{eq:KKT-MPEC.4} to obtain:
\begin{align}
\label{eq8}
2ac_i (D_i+a\bar \lambda_i -\bar b_i) + a d_i -2a \bar \lambda_i + \bar b_i + aI \bar \delta & \nonumber\\
+ a \sum \limits_{j=1}^I \sum \limits_{l=1}^L \pi_{jl}\bar \gamma_l^{-} - a \sum \limits_{j=1}^I \sum \limits_{l=1}^L \pi_{jl}\bar \gamma_l^{+} & = 0
\end{align}
For RNLPs of prosumers $j \in \mathcal{I} \backslash\{i\}$, constraint \eqref{eq:KKT-MPEC.3} holds for $i$, and hence \eqref{eq:KKT-MPEC.2} minus \eqref{eq:KKT-MPEC.3} leads to:
\begin{align}\label{eq8-other}
    2ac_i(D_i+a\bar \lambda_i-\bar b_i)+ad_i -2a\bar \lambda_i + \bar b_i =0
\end{align}
Combining \eqref{eq8}--\eqref{eq8-other}, we have
\begin{align}
    I\bar \delta + \sum \limits_{i=1}^I \sum \limits_{l=1}^L \pi_{il} (\bar \gamma_l^{-}-\bar \gamma_l^{+})=0
\end{align}
\eqref{eq:KKT-MPEC.1}$\times aI +$\eqref{eq8} and \eqref{eq8-other} with $\bar p_i \!=\! D_i\!+\! a\bar \lambda_i\!-\!\bar b_i$ lead to:
\begin{eqnarray}
 &&(2c_i \bar p_i+d_i)-\frac{D_i- \bar p_i}{a(I-1)}-\bar \lambda_i+ \frac{I}{I-1} \sum \limits_{l=1}^L \pi_{il} (\bar \gamma_l^{-}-\bar \gamma_l^{+}) \nonumber  \\ 
 &=& -  \frac{I}{I-1}\bar \delta,\quad \forall i \in \mathcal{I}  \label{eq9}
\end{eqnarray}
where the right-hand side is independent of $i$.
Substitute \eqref{eq:KKT-MPEC.14} into \eqref{eq9} we have for all $i \in \mathcal{I}$:
\begin{align}
     (2c_i \bar p_i+d_i)-\frac{D_i - \bar p_i}{a(I-1)} +\frac{a}{2}\bar \eta & \nonumber\\
    +  \sum \limits_{l=1}^L \pi_{il} (\frac{I}{I-1} \bar \gamma_l^{-}+\frac{a}{2}\bar \alpha_l^{-} -\frac{I}{I-1} \bar \gamma_l^{+}-\frac{a}{2}\bar \alpha_l^{+}) & = -\frac{I}{I-1}\bar \delta \nonumber
\end{align}
Let $\kappa=\frac{a}{2}\bar \eta+\frac{I}{I-1}\bar \delta$ and $\tau_l^{-}=\frac{I}{I-1} \bar \gamma_l^{-}+\frac{a}{2}\bar \alpha_l^{-}$, $\tau_l^{+}=\frac{I}{I-1} \bar \gamma_l^{+}+\frac{a}{2}\bar \alpha_l^{+}$ for all $l\in\mathcal{L}$. One can verify that $(\bar p, \kappa, \tau^{\pm})$ satisfies KKT condition \eqref{eq:central-KKT}, which under A1 implies $\bar p$ is the unique optimal solution of problem \eqref{eq:central}.

Furthermore, by \eqref{eq:KKT-MPEC.2}$+a\times$\eqref{eq:KKT-MPEC.1}, we have for all $i\in\mathcal{I}$:
\begin{align}
\label{eq:eq13}
    \bar \epsilon_i = -\frac{D_i-\bar p_i}{2} 
\end{align}
Combining \eqref{eq:eq13}, \eqref{eq:KKT-MPEC.1}, \eqref{eq:KKT-MPEC.3}, we have for all $i\in\mathcal{I}$:
\bsq
\begin{align}
    \bar \lambda_i ~& = 2c_i\bar p_i+d_i+ \frac{D_i-\bar p_i}{a} \\
    \bar b_i ~& = 2ac_i\bar p_i+ad_i+2(D_i-\bar p_i)
\end{align}
\esq
Therefore, uniqueness of $\bar p$ implies uniqueness of VE $(\bar p,\bar b)$.

\renewcommand{\theequation}{C.\arabic{equation}}
\renewcommand{\thetheorem}{C.\arabic{theorem}}
\section{Analysis of Example 2: Three prosumers}
\label{apen-example}
Under the settings in Example 2, problem \eqref{eq:central} becomes:
\bsq
\label{eq:central-3bus}
\begin{align}
    \mathop{\min}_{p_1,p_2,p_3}~ & \sum \limits_{i=1}^3 \left(cp_i^2+\frac{(D_i-p_i)^2}{4}\right) \\
    \mbox{s.t.}~ & p_1+p_2+p_3 =D_1+D_2+D_3 \\
    ~ & -F \le D_1-p_1 \le F
\end{align}
\esq

The unique optimal solution of \eqref{eq:central-3bus} is:

$\bullet$ 
If $\frac{8c D_1 - 4c D_2 - 4c D_3}{12c+3}  \in (-F,F)$ then
\begin{eqnarray}
    p_i^* & =& \frac{D_i}{4c+1}+\frac{4c\sum_{j=1}^3 D_j}{12c+3},\quad \forall i=1,2,3 \nonumber
\end{eqnarray}

$\bullet$ If $\frac{8c D_1 - 4c D_2 - 4c D_3}{12c+3} \in (-\infty,-F]$ then
\begin{eqnarray}
    p_1^* & = & D_1+F \nonumber\\
    p_2^*& = & \frac{(2c+1)D_2+2c D_3}{4c+1} - \frac{F}{2} \nonumber\\
    p_3^* & =& \frac{(2c+1)D_3+2c D_2}{4c+1} - \frac{F}{2} \nonumber
\end{eqnarray}

$\bullet$ If $\frac{8c D_1 - 4c D_2 - 4c D_3}{12c+3} \in [F,+\infty)$ then
\begin{eqnarray}
    p_1^* & = & D_1-F \nonumber\\
    p_2^*& = & \frac{(2c+1)D_2+2c D_3}{4c+1} + \frac{F}{2} \nonumber\\
    p_3^* & =& \frac{(2c+1)D_3+2c D_2}{4c+1} + \frac{F}{2} \nonumber
\end{eqnarray}


Given bids $b=(b_1, b_2, b_3)$, the prices solved by \eqref{eq:new-rule} are:
\begin{eqnarray}
&&(\lambda _1,\lambda_2,\lambda_3) = \begin{cases} 
\left(\frac{b_1\!+\!b_2\!+\!b_3}{3},  \frac{b_1\!+\!b_2\!+\!b_3}{3}, \frac{b_1\!+\!b_2\!+\!b_3}{3}\right), \forall b\in \mathcal{B}_M \\
\left(b_1\!+\!F,  \frac{b_2\!+\!b_3\!-\!F}{2}, \frac{b_2\!+\!b_3\!-\!F}{2}\right), \forall b\in \mathcal{B}_L \\
\left(b_1\!-\!F,  \frac{b_2\!+\!b_3\!+\!F}{2}, \frac{b_2\!+\!b_3\!+\!F}{2}\right), \forall b\in \mathcal{B}_U
\end{cases} \nonumber
\end{eqnarray}
where
\begin{eqnarray}
&&\mathcal{B}_M := \left\{ b\in \mathbb{R}^3~|~\frac{b_2 \!+\! b_3\! -\!3F}{2}<b_1<\frac{b_2 \!+\! b_3 \!+\! 3F}{2} \right\} \nonumber \\
&&\mathcal{B}_L := \left\{ b\in \mathbb{R}^3~|~ b_1\leq \frac{b_2 + b_3 - 3F}{2} \right\} \nonumber \\
&&\mathcal{B}_U := \left\{ b\in \mathbb{R}^3~|~b_1\geq \frac{b_2 + b_3 + 3F}{2} \right\} \nonumber 
\end{eqnarray}
are mutually exclusive sets whose union is $\mathbb{R}^3$.

Given $(b_2, b_3)$, it can be verified that prosumer 1's objective function is continuous on $b_1 \in \mathbb{R}$; moreover, it is linear and strictly decreasing on $(-\infty,~(b_2 \!+\! b_3\! -\!3F)/2]$, quadratic and strictly convex on $((b_2 \!+\! b_3\! -\!3F)/2, ~(b_2 \!+\! b_3\! +\!3F)/2)$, and linear and strictly increasing on $[(b_2 \!+\! b_3\! +\!3F)/2, ~+\infty)$. 
However, given $(b_1, b_3)$, the structure of prosumer 2's objective is more complicated. It is continuous on $b_2 \in \mathbb{R}$, and is quadratic and strictly convex \emph{piecewise} across the three segments divided by $(2b_1 -b_3 \pm 3F)$. Given $(b_1, b_2)$, the structure of prosumer 3's objective function is similar to prosumer 2. Therefore, when analyzing the three-prosumer version of game $\mathcal{G}$ in \eqref{eq:sharing-game}, we not only pay attention to the relationship between the axis of symmetry and boundary points of every quadratic segment, but also screen all the possible local optima to identify each prosumer's globally optimal response.        

Next is the detailed analysis in different subsets of $b\in\mathbb{R}^3$.       

\emph{Part 1: Analysis of GNE in $\mathcal{B}_M$.} Suppose there is a GNE $b \in \mathcal{B}_M$,\footnote{We refer to $b$ as GNE since $(\lambda, q, p)$ can be uniquely determined by $b$.} then it must satisfy (as a necessary condition):   
\begin{eqnarray}
&&b_i = \frac{12cD_i +(4c-1)\sum_{j\neq i} b_j}{8c+4},~\forall i=1,2,3 \label{eq:cond:B_M_1} \\
&& \frac{b_2 \!+\! b_3\! -\!3F}{2} < b_1 <\frac{b_2 \!+\! b_3\! +\!3F}{2} \label{eq:cond:B_M_2}
\end{eqnarray}
where \eqref{eq:cond:B_M_1} is the axis of symmetry of the central quadratic segment in each prosumer's objective function given other prosumers' bids. Solving \eqref{eq:cond:B_M_1}, we obtain the only possible GNE \emph{candidate} in $\mathcal{B}_M$:  
\begin{eqnarray}
 b_i  = \frac{4cD_i}{4c+1} +\frac{8c^2-2c}{12c+3}\sum_{j=1}^3 D_j, ~\forall i =1,2,3 \label{eq:GNE:B_M}
\end{eqnarray}
The GNE candidate $b$ in \eqref{eq:GNE:B_M} satisfies \eqref{eq:cond:B_M_2} if and only if 
\begin{eqnarray} 
    \frac{8c D_1 - 4c D_2 - 4c D_3}{12c+3} \in (-F,F) \label{eq:cond:B_M:parametric}
\end{eqnarray}
which is the condition for both lower and upper line flow constraints to be inactive at the unique optimal solution of \eqref{eq:central-3bus}. If this condition holds, the power profile $p$ determined by $b$ in \eqref{eq:GNE:B_M} is indeed the unique optimal solution of \eqref{eq:central-3bus}. 

The analysis so far reveals uniqueness and optimality of GNE in $\mathcal{B}_M$ \emph{if} one exists. However, even if \eqref{eq:cond:B_M:parametric} holds, the only GNE candidate in \eqref{eq:GNE:B_M} may still be disqualified for GNE, in which case no GNE exists in $\mathcal{B}_M$. Indeed, condition \eqref{eq:cond:B_M_1} $\&$ \eqref{eq:cond:B_M_2}, equivalently \eqref{eq:GNE:B_M} $\&$ \eqref{eq:cond:B_M:parametric}, is necessary but not sufficient for $b$ to be a GNE in $\mathcal{B}_M$. It is still possible that given $(b_1,b_3)$ in \eqref{eq:GNE:B_M}, prosumer 2 only attains a local minimum at $b_2$ over its central quadratic segment $(2b_1 \!-\!b_3 \!-\! 3F,~2b_1\! -\!b_3 \!+\! 3F)$, whereas another segment contains a local minimum that is even lower (better) than $b_2$. A similar scenario might also happen to prosumer 3. Such a scenario disqualifies \eqref{eq:GNE:B_M} for GNE and causes nonexistence of GNE in $\mathcal{B}_M$. 

Derivation of an analytic condition for existence of GNE in $\mathcal{B}_M$ is tedious and does not add much insight. Therefore, we end \emph{Part 1} simply by providing two numerical examples: 
\begin{itemize}
\item $c=1$, $D_1 = D_2 = 1$, $D_3 =0$, $F = 0.3$. We obtain $b_1=b_2=1.6$, $b_3=0.8$ by \eqref{eq:GNE:B_M}, and verify that it is the unique GNE in $\mathcal{B}_M$ since $b_i$ is the global optimum of every prosumer $i$'s objective given $b_{-i}$. We further determine $p_1=p_2=0.7333$, $p_3 = 0.5333$ which is the unique optimal of \eqref{eq:central-3bus} where no congestion occurs.
\item All the parameters, including the GNE candidate $(p,b)$, are the same as above, except $F=0.27$. Given $(b_1,b_3)$, prosumer 2 attains a second local minimum $b_2' = 1.535 < 2b_1 \!-\! b_3\! -\!3F$, at which $\lambda_2' = \frac{b_2'+b_3 +F}{2} = 1.3025$, $p_2'=0.7675$, and prosumer 2's objective $\Gamma_2(b_2') = 0.8919<0.8933=\Gamma_2(b_2)$. Therefore, the only GNE candidate $(p,b)$ is disqualified and no GNE exists in $\mathcal{B}_M$. 
\end{itemize}  

\emph{Part 2: Analysis of GNE(s) in $\mathcal{B}_L$.} Suppose there is a GNE $b \in \mathcal{B}_L$, then it must satisfy (as a necessary condition): 
\begin{eqnarray}
b_1 &=& \frac{b_2 + b_3 - 3F}{2},\quad\textnormal{and} \label{eq:cond:B_L_1}
\end{eqnarray}
\bsq
\label{eq:cond:B_L_2}
\begin{align}
&&\frac{12cD_1 +(4c-1)(b_2 +b_3)}{8c+4} \leq \frac{b_2 + b_3 - 3F}{2} \label{eq:cond:B_L_2:1} \\
&&\frac{12cD_2 +(4c-1)(b_1 +b_3)}{8c+4} \geq 2b_1-b_3+3F \label{eq:cond:B_L_2:2} \\
&&\frac{12cD_3 +(4c-1)(b_1 +b_2)}{8c+4} \geq 2b_1-b_2+3F \label{eq:cond:B_L_2:3} \\
&&\frac{2c}{c+1}(D_2+\frac{b_3-F}{2}) \leq 2b_1-b_3+3F \label{eq:cond:B_L_2:4} \\
&&\frac{2c}{c+1}(D_3+\frac{b_2-F}{2}) \leq 2b_1-b_2+3F \label{eq:cond:B_L_2:5}
\end{align}
\esq
where \eqref{eq:cond:B_L_1} must hold because prosumer 1's objective is strictly deceasing on $b_1 \in (-\infty, ~\frac{b_2 + b_3 - 3F}{2}]$. The left-hand-sides of \eqref{eq:cond:B_L_2:4} and \eqref{eq:cond:B_L_2:5} are, respectively, the axes of symmetry of the right quadratic segments of prosumers 2 and 3's objectives. Indeed, if any $b$ satisfying \eqref{eq:cond:B_L_1} and \eqref{eq:cond:B_L_2} also satisfies the following inequalities, then it suffices for $b$ to be a GNE:\footnote{This does not imply \eqref{eq:cond:B_L_1}--\eqref{eq:cond:B_L_3} is necessary and sufficient for GNE in $\mathcal{B}_L$, since $b$ may satisfy \eqref{eq:cond:B_L_1}--\eqref{eq:cond:B_L_2}, violate \eqref{eq:cond:B_L_3}, and still be a GNE.}   
\bsq
\label{eq:cond:B_L_3}
\begin{align}
&&\frac{2c}{c+1}(D_2+\frac{b_3+F}{2}) \geq 2b_1-b_3-3F \label{eq:cond:B_L_3:4} \\
&&\frac{2c}{c+1}(D_3+\frac{b_2+F}{2}) \geq 2b_1-b_2-3F \label{eq:cond:B_L_3:5}
\end{align}
\esq
where the left-hand-sides are, respectively, axes of symmetry of left quadratic segments of prosumers 2 and 3's objectives. 

Condition \eqref{eq:cond:B_L_1}--\eqref{eq:cond:B_L_2:3} implies 
\begin{eqnarray} 
    \frac{8c D_1 - 4c D_2 - 4c D_3}{12c+3} \in (-\infty,-F] \label{eq:cond:B_L:parametric}
\end{eqnarray}
i.e., the lower line flow constraint is reached at the unique optimal solution of \eqref{eq:central-3bus}. Considering a necessary condition $\frac{8c D_1 - 4c D_2 - 4c D_3}{12c+3} \in [F, +\infty)$ which can be derived in a similar way for any GNE in $\mathcal{B}_U$, as well as \eqref{eq:cond:B_M:parametric}, we conclude that given any network parameters, the three-prosumer version of game \eqref{eq:sharing-game} has GNE(s) in \emph{at most one} of $\mathcal{B}_M$, $\mathcal{B}_L$, and $\mathcal{B}_U$. 

Replace $b_1$ in \eqref{eq:cond:B_L_2} and \eqref{eq:cond:B_L_3} with \eqref{eq:cond:B_L_1}, so that we can focus on the $(b_2,b_3)$ space. In general, if \eqref{eq:cond:B_L:parametric} holds, game \eqref{eq:sharing-game} may have no GNE in $\mathcal{B}_L$ (e.g., if no $(b_2,b_3)$ satisfies \eqref{eq:cond:B_L_2}) or uncountably many GNEs in $\mathcal{B}_L$ (e.g., if there exists a polygon on the $(b_2, b_3)$ space defined by \eqref{eq:cond:B_L_2} and \eqref{eq:cond:B_L_3}). We provide the following numerical examples to illustrate these cases. 

\begin{itemize}
\item $c=1$, $D_1 =0$,  $D_2 = D_3 =1$, $F = 1/3$. In this case, \eqref{eq:cond:B_L_2:2}--\eqref{eq:cond:B_L_2:5} define a polygon on the $(b_2, b_3)$ space, which is contained in the triangle defined by \eqref{eq:cond:B_L_2:1} and \eqref{eq:cond:B_L_3}. Therefore, all the points in this polygon are GNEs.
Some of these GNEs have the same power profile $p$ which is the unique optimal solution of \eqref{eq:central-3bus} where the lower line flow constraint is binding, while others do not. For instance, $b=(b_1,b_2,b_3)=(1.18, 1.68, 1.68)$ and $b=(1.22, 1.72, 1.72)$ are two GNEs which lead to the same $p =(p_1,p_2,p_3)= (\frac{1}{3}, \frac{5}{6}, \frac{5}{6})$ that is optimal for \eqref{eq:central-3bus}, whereas $b=(1.21, 1.7, 1.72)$ is a GNE that leads to $p=(\frac{1}{3}, \frac{253}{300}, \frac{247}{300})$ which is not optimal for \eqref{eq:central-3bus}. 

\item $c=1$, $D_1 =0$,  $D_2 = 1$, $D_3 =2$, $F = 1/3$. Even though \eqref{eq:cond:B_L:parametric} is satisfied and \eqref{eq:central-3bus} has a unique optimal solution where the lower line flow constraint is binding, there is no point on the $(b_2, b_3)$ space that satisfies \eqref{eq:cond:B_L_2:2}--\eqref{eq:cond:B_L_2:5} simultaneously. In this case, game \eqref{eq:sharing-game} has no GNE.    
\end{itemize}  

We skip the analysis for $\mathcal{B}_U$ due to its similarity to \emph{Part 2}.

\renewcommand{\theequation}{D.\arabic{equation}}
\renewcommand{\thetheorem}{D.\arabic{theorem}}
\section{Two lemmas}
\label{apen-2}

We assume condition A1 holds throughout Appendix \ref{apen-2}. Recall $q_i =-a\lambda_i +b_i,\forall i \in \mathcal{I}$. Then the networked energy sharing game $\mathcal{G}$ can be equivalently written as follows.

\noindent Decision-making of every prosumer $i \in \mathcal{I}$:
\bsq
\label{eq:sharing-upper-m}
\bq
\mathop {\min} \limits_{p_i,b_i} && c_i (p_i)^2+d_i p_i +\frac{1}{a}q_i(b)(-q_i(b)+b_i) \label{eq:sharing-upper-m.1}\\
\rm{s.t.}&& p_i+q_i(b)=D_i \label{eq:sharing-upper-m.2}
\eq
\esq
Market platform's problem:
\bsq
\label{eq:sharing-lower-m}
\bq
\mathop {\min} \limits_{q_i,\forall i \in \mathcal{I}} && \sum_{i=1}^I (q_i-b_i)^2 \label{eq:sharing-lower-m.1}\\
\rm{s.t.}&& \sum_{i=1}^I q_i=0 \label{eq:sharing-lower-m.2}\\
&& -\tilde F_l \le \sum_{i=1}^I \pi_{il} q_i \le F_l,~\forall l \in \mathcal{L}  \label{eq:sharing-lower-m.3}
\eq
\esq
 where problem \eqref{eq:sharing-lower-m} is parameterized by $b=(b_i,~\forall i \in \mathcal{I})$. Due to feasibility of \eqref{eq:sharing-lower-m} by condition A1 and strong convexity of objective \eqref{eq:sharing-lower-m.1}, problem \eqref{eq:sharing-lower-m} attains a unique optimal solution $q(b)$ as a function of $b$. The $i$-th element of $q(b)$ is $q_i(b)$ in \eqref{eq:sharing-upper-m}.  
We sometimes fix $b_{-i}:=(b_j, ~\forall j\in\mathcal{I}, j\neq i)$ and consider $b_i$ as the only varying parameter. In that case, the optimal solution of \eqref{eq:sharing-lower-m} is a function of $b_i$ only, denoted as $q(b_i)$, when the slight abuse of notation $q(b)$ versus $q(b_i)$ would not cause confusion.
We sometimes also write $q(b)$ or $q(b_i)$ as $q^*$ to skip the argument $b$ or $b_i$ for conciseness.


\begin{lemma}
\label{lemma-2}
The optimal solution $q(b)$ of \eqref{eq:sharing-lower-m} is continuous and piece-wise linear in $b$. Moreover, the domain $\mathbb{R}^I$ of $b$ is composed of \emph{finitely} many polyhedra, within each $q(b)$ is linear in $b$ and satisfies:
\bq\label{eq:lemma-2}
0 \le \frac{\partial q_i(b)}{\partial b_i} \le \frac{I-1}{I}, \quad \forall i \in \mathcal{I}
\eq
\end{lemma}
\begin{proof}
For problem \eqref{eq:sharing-lower-m} with quadratic objective and affine constraints, Theorem 6.11 in \cite{still2018lectures}  verifies continuity and piece-wise linearity of $q(b)$ in $b$, whose domain $\mathbb{R}^I$ is composed of finitely many polyhedra, on each $q(b)$ is linear.

It is sufficient to prove \eqref{eq:lemma-2} within an arbitrary polyhedron of $b$, for an arbitrary $i\in\mathcal{I}$. 
Let $db:=[...,0,db_i,0,...]^T \in \mathbb{R}^I$ be a vector of all zeros except $i$-th element $db_i$. As $b$ changes to $b+db$, the optimal $q(b)=q^*$ changes to $q(b+db)=q^*+d q^*$. 

Since $q(b)$ is linear in $b$, we can prove $\partial q_i(b) / \partial b_i \geq 0$ by showing $dq_i^* \geq 0$ for any sufficiently small $db_i>0$ that retains $b+db$ in the same polyhedron with $b$. To this end, we assume $dq_i^*<0$ for some $db_i>0$ and shall deduce a contradiction. Denote the objective \eqref{eq:sharing-lower-m.1} as $g(q,b)$. We then have:
\begin{align}
~& g(q^*,b)+g(q^*+d q^*, b+d b) \nonumber\\
=~ & (q_i^*-b_i)^2+\sum \limits_{j\in\mathcal{I}, ~j \ne i}(q_j^*-b_j)^2+ (q_i^*+d q_i^*-b_i-d b_i)^2 \nonumber\\
~ & +\sum \limits_{j\in\mathcal{I}, ~j \ne i}(q_j^*+d q_j^*-b_j)^2 \nonumber
\end{align}
and
\begin{align}
~ & g(q^*+\frac{1}{2}d q^*,b)+g(q^*+\frac{1}{2}d q^*,b+d b) \nonumber\\
=~ & (q_i^*+\frac{1}{2}d q_i^*-b_i)^2+\sum \limits_{j\in\mathcal{I}, ~j \ne i} (q_j^*+\frac{1}{2}d q_j^*-b_j)^2 \nonumber\\
~ & + (q_i^*+\frac{1}{2}d q_i^*-b_i- d b_i)^2+\sum \limits_{j\in\mathcal{I},j \ne i} (q_j^*+\frac{1}{2}d q_j^*-b_j)^2 \nonumber
\end{align}
Apply the following two inequalities to the two above:
\begin{eqnarray}
\frac{(q_j^*-b_j)^2+(q_j^*+d q_j^*-b_j)^2}{2} &\ge & (q_j^*+\frac{1}{2}d q_j^*-b_j)^2, \nonumber
\\&& \qquad  \forall j \in \mathcal{I},~j\neq i \nonumber
\end{eqnarray}
and
\bq
&& (q_i^*-b_i)^2+(q_i^*+d q_i^*-b_i-d b_i)^2-(q_i^*+\frac{1}{2}d q_i^*-b_i)^2 \nonumber\\
&& -(q_i^*+\frac{1}{2}d q_i^*-b_i-d b_i)^2 = \frac{(d q_i^*)^2}{2}-d q_i^* d b_i >0 \nonumber
\eq
The result is
\bq
&& g(q^*,b)+g(q^*+d q^*, b+d b) \nonumber\\
&> & g(q^*+\frac{1}{2}d q^*,b)+g(q^*+\frac{1}{2}d q^*,b+d b) \nonumber
\eq
Convexity of \eqref{eq:sharing-lower-m.2}--\eqref{eq:sharing-lower-m.3} implies feasibility of $q^*+\frac{1}{2}d q^*$, so that the equation above contradicts the fact that $q^* $ and $q^*+d q^*$ are the optimal solutions of  \eqref{eq:sharing-lower-m} corresponding to $b$ and $b+db$, respectively. Therefore, we have proved $\partial q_i(b) / \partial b_i \geq 0$.

We now prove $\partial q_i(b) / \partial b_i \leq (I-1)/I$. Similar to the preceding proof, we assume $d q_i^*/d b_i > (I-1)/I$ for some sufficiently small $db_i > 0$ that retains $b+db$ in the same polyhedron with $b$, and deduce a contradiction below. 
Indeed, $q^*$ and $q^*+dq^*$ are the projections of $b$ and $b+db$, respectively, onto affine subspace $\mathcal{A}$ composed of \eqref{eq:sharing-lower-m.2} plus the \emph{same} set of binding inequalities from \eqref{eq:sharing-lower-m.3}. 
Note that $dq^*$ lies in the linear subspace parallel to $\mathcal{A}$, which implies $(q^*-b) \perp d q^*$ and $(q^*+d q^*-b-d b) \perp d q^*$. 
Also note that $\{q^* + \alpha dq^* \ | \ \alpha \in \mathbb{R}\}$ is a subset of $\mathcal{A}$ and an affine subspace itself, denoted as $\mathcal{A}'$, and $q^*$ and $q^*+dq^*$ are, respectively, the projections of $b$ and $b+db$ onto $\mathcal{A}'$ as well. We define vector
\begin{eqnarray} 
\tilde q &:=&(q_1^*-\frac{1}{I}d b_i,...,q_i^*+\frac{I-1}{I}d b_i,,...,q_I^*-\frac{1}{I}d b_i) \nonumber
\end{eqnarray}
and denote the projection of $\tilde q$ onto $\mathcal{A}'$ as 
\begin{eqnarray}
\check q = q^* + \check \alpha dq^* \nonumber
\end{eqnarray} 
which satisfies $(\check q-\tilde{q}) \perp d q^*$, i.e., the inner product satisfies: 
\begin{eqnarray}
& &(\check\alpha d q_1^*+\frac{1}{I}d b_i,...,\check\alpha d q_i^*-\frac{I-1}{I}d b_i,,...,\check\alpha d q_I^*+\frac{1}{I}d b_i) \nonumber\\
&&\cdot (d q_1^*,..., d q_i^*,...,d q_I^*) =0 \label{eq:proof-lemma2:1}
\end{eqnarray}
Applying $\sum_{i=1}^I dq_i^*=0$ to \eqref{eq:proof-lemma2:1}, we have
\begin{eqnarray}
    0< \check \alpha & =  & \frac{d b_i d q_i^*}{\sum_{j=1}^I (d q_j^*)^2} 
    \le  \frac{d b_i d q_i^*}{(d q_i^*)^2+\frac{1}{I-1}(d q_i^*)^2} \nonumber\\
    & =  & \frac{(I-1)/I}{d q_i^*/db_i} <1 \nonumber
\end{eqnarray}
Moreover, by $(q^*-b)\perp d q^*$ and $\sum_{i=1}^{I} d q_i^*=0$, we have
\begin{eqnarray}
&& (b+d b-\tilde{q}) \cdot d q^* \nonumber\\
&=&  (b_1\!-\!q_1^*\!+\!\frac{d b_i}{I},...,b_I\!-\!q_I^*\!+\!\frac{d b_i}{I})  \cdot (d q_1^*,...,d q_I^*) = 0~ \IEEEeqnarraynumspace \label{eq:proof-lemma2:2}
\end{eqnarray}
Equation \eqref{eq:proof-lemma2:2} and $(\check{q}-\tilde{q}) \perp dq^*=0$ in the preceding texts imply $(\check{q}-b-d b) \perp d q^*$. The unique projection of $b+db$ onto $\mathcal{A}'$ is thus $\check q = q^*+ dq^*$, which contradicts $\check \alpha <1$.
\end{proof}

\begin{lemma}
\label{lemma-4}
 Let $\lambda(b)$ denote the optimal solution of market-clearing problem \eqref{eq:new-rule} parameterized by $b$. Then $(\bar p,\bar b)$ is a GNE of the improved energy sharing paradigm \eqref{eq:modified-obj} and $\bar \lambda^r$ is the regulated price at $(\bar p,\bar b)$, if and only if \eqref{eq:cond:improved-GNE} holds.
\end{lemma}
\begin{proof}
Subject to \eqref{eq:modified-obj.2}, the term $u_i(b,p_i)$ in \eqref{eq:modified-obj.1} equals:
\begin{eqnarray}
\mathop{\max}\left\{\frac{\!-\!q_i(b)\!+\!b_i}{a} q_i(b),\left(2c_i \left(D_i \!-\! q_i(b)\right) \!+\! d_i \!-\! \frac{q_i(b)}{a(I\!-\!1)}\right) q_i(b)\right\} \nonumber
\end{eqnarray}
where $q_i(b)=-a \lambda_i(b) + b_i$, $\forall i\in\mathcal{I}$ is the optimal solution of \eqref{eq:sharing-lower-m} which is equivalent to \eqref{eq:new-rule}.

We consider an arbitrary vector $b_{-i}$ and fix it. Lemma \ref{lemma-2} implies that 
$(-q_i(b)+b_i)/a$ is \emph{strictly} monotonically increasing on $b_i \in \mathbb{R}$ with image $(-\infty,\infty)$ and 
$\left(2c_i \left(D_i \!-\! q_i(b)\right) \!+\! d_i \!-\! \frac{q_i(b)}{a(I\!-\!1)}\right)$ is monotonically decreasing on $b_i \in \mathbb{R}$. 
Hence there exists a \emph{unique} $b_i^* \in \mathbb{R}$ at which these two terms are equal, with
$(-q_i(b)+b_i)/a>\left(2c_i \left(D_i \!-\! q_i(b)\right) \!+\! d_i \!-\! \frac{q_i(b)}{a(I\!-\!1)}\right)$ if $b_i >b_i^*$ and vice versa. Note that $b_i^*$ depends on $b_{-i}$.  

Consider the case $q_i(b_i^*)>0$ with the fixed $b_{-i}$ in the last paragraph. For $b_i>b_i^*$, there is always $q_i(b)>0$, and the modified objective \eqref{eq:modified-obj.1} of prosumer $i\in\mathcal{I}$ becomes:
\begin{eqnarray}
    \tilde \Gamma_i =  c_i\left(D_i\!-\!q_i(b)\right)^2 + d_i \left(D_i\!-\!q_i(b)\right) + \frac{\!-\!q_i(b)\!+\!b_i}{a}q_i(b) \label{eq:proof-lemma-4:1}
\end{eqnarray}
whose derivative over $b_i$ is
\begin{align}
    \frac{d {\tilde \Gamma}_i}{d b_i} = ~& \left(-2c_i(D_i-q_i) -d_i -\frac{2}{a}q_i+\frac{b_i}{a} \right)   \frac{dq_i}{d b_i}+\frac{q_i}{a} \nonumber\\
    \ge ~ & -\left(\frac{q_i}{a(I-1)}+\frac{q_i}{a}\right)\frac{d q_i}{d b_i}+\frac{q_i}{a} \nonumber\\
    \ge ~ & -\frac{q_i}{a}+\frac{q_i}{a} = 0 \nonumber
\end{align}
where the inequalities are due to $0 \leq d q_i/db_i \leq (I-1)/I$ by Lemma \ref{lemma-2}, which cannot simultaneously attain equality and thus renders $d {\tilde \Gamma}_i/db_i$ \emph{strictly} positive.

For $b_i \leq b_i^*$, there exists $\hat b_i \in [-\infty,  b_i^*)$ such that $q_i(b_i^*)\geq q_i(b)\geq 0$ for $b_i \in \left[\hat b_i, b_i^*\right]$ and $q_i(b) < 0$ for $b_i \in (-\infty, \hat b_i)$. Let $\hat b_i=-\infty$ if $q_i(b) \geq 0$ for all $b_i \in \mathbb{R}$. If $b_i \in \left[\hat b_i, b_i^*\right]$, we have:
\begin{eqnarray}
    \tilde \Gamma_i   &= &  c_i\left(D_i-q_i(b)\right)^2+d_i\left(D_i-q_i(b)\right) \nonumber\\
    & & + \left(2 c_i\left(D_i-q_i(b)\right)+d_i-\frac{q_i(b)}{a(I-1)}\right)q_i(b) \nonumber
\end{eqnarray}
whose derivative over $b_i$ is
\begin{eqnarray}
    \frac{d\tilde \Gamma_i}{db_i} = 
     \left(-2c_i -\frac{2}{a(I-1)}\right) q_i \frac{dq_i}{db_i} \le 0 \nonumber
\end{eqnarray}
Note that for an arbitrary $\epsilon > 0$ that is sufficiently small, to maintain $d{\tilde \Gamma}_i/db_i\equiv 0$ 
for $b_i \in (b_i^*-\epsilon, b_i^*]$, there must be $dq_i/db_i \equiv 0$ in the same region, because $q_i(b) > 0$ in this region due to continuity of $q_i$ at $b_i^*$.

If $b_i \in (-\infty, \hat b_i)$, we have the same $\tilde \Gamma_i$ as \eqref{eq:proof-lemma-4:1}, but $d\tilde \Gamma_i/db_i < 0$ (strictly). 
To summarize, $\tilde \Gamma_i$ is a function of $b_i$ strictly decreasing on $(-\infty,~\hat b_i)$, decreasing on $[\hat b_i,~b_i^*]$, and strictly increasing on $(b_i^*,~+\infty)$.
Therefore, $b_i^*$ is a minimizer of $\tilde \Gamma_i$ but might not be the unique one. In general, all the minimizers of $\tilde \Gamma_i$ form a set $[b_i^*-\epsilon,~b_i^*]$ for some $\epsilon \geq 0$, on which $q_i$ keeps unchanged; in this case, we assume that prosumer $i$ shall just return $b_i^*$ as its bid.
A similar analysis for the case $q_i(b_i^*)\leq 0$ leads to the same result that prosumer $i$ returns $b_i^*$. Since $b_i^*$ depends on $b_{-i}$, we write it as $b_i^*(b_{-i})$.

Therefore, a point $(\bar p, \bar b)$ is a GNE of the improved energy sharing paradigm \eqref{eq:modified-obj}, if and only if $(\bar p, \bar b)$ satisfies \eqref{eq:modified-obj.2} and every prosumer $i\in\mathcal{I}$ returns $\bar b_i = b_i^*(\bar b_{-i})$, i.e.,
\begin{eqnarray}
\frac{-q_i(\bar b)+\bar b_i}{a} =  2c_i \left(D_i - q_i(\bar b)\right) + d_i - \frac{q_i(\bar b)}{a(I-1)},~\forall i \in\mathcal{I}  \label{eq:cond:b-at-GNE}
\end{eqnarray}
The combination of \eqref{eq:cond:b-at-GNE}, \eqref{eq:modified-obj.2}, and price regulation rule \eqref{eq:priceregulate} is equivalent to \eqref{eq:cond:improved-GNE}, which completes the proof.
\end{proof}

\renewcommand{\theequation}{E.\arabic{equation}}
\renewcommand{\thetheorem}{E.\arabic{theorem}}
\section{Proof of Theorem \ref{Thm:prop-2}}
\label{apen-5}
\emph{Proof for existence of GNE.} Recall that if A1 holds, problem \eqref{eq:central} has a \emph{unique} optimal $\bar p$ and a dual optimal $(\bar \kappa, \bar \tau^{\pm})$ which together satisfy the KKT condition \eqref{eq:central-KKT}. By $\bar \eta = \frac{2}{a} \bar \kappa$, $\bar \alpha^{\pm} = \frac{2}{a} \bar \tau^{\pm}$, and $\bar \lambda_i^r =  2c_i \bar p_i + d_i - \frac{D_i - \bar p_i}{a(I-1)}$, $\bar b_i = D_i - \bar p_i + a \bar \lambda_i^r$ for all $i \in \mathcal{I}$, we construct a point $(\bar \lambda^r; \bar \eta, \bar \alpha^\pm)$ which satisfies the KKT condition \eqref{eq:lowerKKT} and is thus primal-dual optimal for the market-clearing problem \eqref{eq:new-rule} parameterized by $\bar b$, i.e., $\bar \lambda^r = \lambda(\bar b)$. Therefore, $(\bar p, \bar b)$ and $\bar \lambda^r$ satisfy \eqref{eq:cond:improved-GNE} and constitute a GNE of the improved game \eqref{eq:modified-obj} by Lemma \ref{lemma-4}.

\emph{Proof for uniqueness of GNE}. Consider an arbitrary GNE $(\bar p^{\prime},\bar b^{\prime})$ of the improved game \eqref{eq:modified-obj}. By Lemma \ref{lemma-4}, the regulated price ${\bar \lambda}^{r\prime}$ at $(\bar p^{\prime},\bar b^{\prime})$ must satisfy ${\bar \lambda}^{r\prime}_i= \lambda_i(\bar b^{\prime}) = 2 c_i \bar p^{\prime}_i+d_i-\frac{D_i - \bar p^{\prime}_i}{a(I-1)}$ for all $i \in \mathcal{I}$, i.e., $\bar \lambda^{r\prime}$ is the unique optimal solution of \eqref{eq:new-rule} parameterized by $\bar b^{\prime}$. Therefore, there exists $(\bar \eta^{\prime}, \bar \alpha^{\pm\prime})$ which together with $(\bar \lambda^{r\prime}, \bar b^{\prime})$ satisfies the KKT condition \eqref{eq:lowerKKT}. 
Constructing $\bar \kappa^{\prime} = \frac{a}{2} \bar\eta^{\prime}$ and $\bar \tau^{\pm\prime}= \frac{a}{2} \bar\alpha^{\pm\prime}$ and noticing $D_i - \bar p^{\prime}_i = -a {\bar \lambda}^{r\prime}_i + \bar b^{\prime}_i$ by Lemma \ref{lemma-4}, we obtain $(\bar p^{\prime}; \bar \kappa^{\prime}, \bar \tau^{\pm\prime})$ which satisfies \eqref{eq:central-KKT} and is thus a primal-dual optimal of \eqref{eq:central}. Therefore, $\bar p^{\prime} = \bar p$ must be the unique primal optimal of \eqref{eq:central}.
As a result, ${\bar \lambda}^{r\prime}_i = 2 c_i \bar p_i+d_i-\frac{D_i - \bar p_i}{a(I-1)}$ and $\bar b^{\prime}_i = D_i - \bar p_i + a {\bar \lambda}^{r\prime}_i$ for all $i \in \mathcal{I}$ are, respectively, the unique price and bid for GNE.

\renewcommand{\theequation}{F.\arabic{equation}}
\renewcommand{\thetheorem}{F.\arabic{theorem}}
\section{Proof of Proposition \ref{Thm:prop-3}}
\label{apen-4}
Select and fix an arbitrary $i \in \mathcal{I}$. Given other prosumers' strategies $\bar b_{-i}=(\bar b_j,\forall j \neq i)$, let $\lambda_{-i}^* =(\lambda_j^*,\forall j \neq i)$ and $(\eta^{\prime*}, \alpha^{\pm\prime*})$ denote the primal optimal and a dual optimal of the pricing problem over $(I-1)$ prosumers excluding $i$: 
\bsq
\label{eq:sharing-N-1}
\begin{align}
\mathop{\min}\limits_{\lambda_j,j \ne i} ~ & \sum \limits_{j \ne i} \lambda_j^2\\
~& \sum \limits_{j \ne i} (a\lambda_j-\bar b_j)=0:\eta^{\prime}\\
~& -\tilde F_l \!\le\! \sum \limits_{j \ne i} \!\pi_{jl} (-a\lambda_j\!+\!\bar b_j) \!\le\! F_l:\alpha_l^{-\prime},
\!\alpha_l^{+\prime}, \forall l \in \mathcal{L}
\end{align}
\esq
If prosumer $i$ bids $\hat b_i=-\frac{a^2}{2}(\eta^{\prime*}+\sum_{l=1}^L \pi_{il}\alpha_l^{-\prime*}-\sum_{l=1}^L \pi_{il}\alpha_l^{+\prime*})$, by KKT condition \eqref{eq:lowerKKT}, solving problem \eqref{eq:new-rule} for bids $(\hat b_i, \bar b_{-i})$ leads to price $(\hat \lambda_i, \lambda_{-i}^*)$ which satisfies $\hat q_i = -a\hat \lambda_i + \hat b_i = 0$ and $\hat p_i = D_i$. Therefore, $\tilde \Gamma_i(\hat b_i,\bar b_{-i}) = J_i(D_i)$.\footnote{We skip input $p$ for function $\tilde \Gamma_i(\cdot)$ since $p$ is a function of $b$.} By definition of GNE, $\tilde \Gamma_i(\bar p, \bar b) \leq \tilde \Gamma_i(\hat b_i, \bar b_{-i}) = J_i(D_i)$.

\renewcommand{\theequation}{G.\arabic{equation}}
\renewcommand{\thetheorem}{G.\arabic{theorem}}
\section{Proof of Proposition \ref{Thm:prop-5}}
\label{apen-6}

By Proposition \ref{Thm:prop-2}, $\bar p(I)$ at GNE is optimal for \eqref{eq:central}. Define $\Omega(p):=\sum_{i=1}^I (D_i-p_i)^2$. Since problems \eqref{eq:AGG} and \eqref{eq:central} have the same feasible set, comparing their objectives leads to: 
\begin{eqnarray}
J(\tilde p(I)) &\leq& J(\bar p(I)) \nonumber \\
J(\tilde p(I)) + \frac{\Omega(\tilde p(I))}{2a(I-1)} &\geq& J(\bar p(I)) + \frac{\Omega(\bar p(I))}{2a(I-1)}\nonumber 
\end{eqnarray}
By assumption, we have (dropping $I$ for conciseness):
\begin{align}
0 \le 	J(\bar p)-J(\tilde p) \le  \frac{\Omega(\tilde{p})}{2a(I-1)} - \frac{\Omega({\bar p})}{2a(I-1)} \le \frac{C_1 I}{2a(I-1)} \nonumber
\end{align}
and therefore:
\begin{align}
    1 \leq \mbox{PoA}(I)  \!=\! 1+\frac{J(\bar p)-J(\tilde p)}{J(\tilde p)} \leq 1+ \frac{C_1}{2a(I-1)C_2} \nonumber
\end{align}

\renewcommand{\theequation}{H.\arabic{equation}}
\renewcommand{\thetheorem}{H.\arabic{theorem}}
\section{Proof of Proposition \ref{Thm:prop-4}}
\label{apen-7}
By Proposition \ref{Thm:prop-2}, $\bar p$ at GNE is the optimal solution of problem \eqref{eq:central}. Combining the price expression \eqref{eq:cond:improved-GNE.1} and the KKT condition \eqref{eq:central-KKT.1} for \eqref{eq:central}, we immediately obtain \eqref{eq:price-structure}.

\renewcommand{\theequation}{I.\arabic{equation}}
\renewcommand{\thetheorem}{I.\arabic{theorem}}
\section{Convergence of \textbf{Algorithm 1}} 
\label{apen-8}
Given $b^k$, the platform's update is equivalent to
\bsq
\label{eq:platform-update}
\begin{align}
q^{k+1}= ~&\mbox{argmin}\{\theta_1(q)-\frac{(b^k)^Tq}{a} \nonumber\\
& +\frac{||q-q^k+b^{k-1}-b^k||^2}{2a} | ~q \in \mathcal{Q}\} \label{eq:platform-update.1}\\
\lambda^{k+1}= ~ & \frac{1}{a}(b^k-q^{k+1}) \label{eq:platform-update.2}
\end{align}
\esq
where $\theta_1(q):=\frac{1}{2a}\sum \limits_{i=1}^I q_i^2$ and 
$$\mathcal{Q}:=\left\{q~|~\mbox{s.t.}~\sum \limits_{i=1}^I q_i = 0, -\tilde F_l \le \sum \limits_{i=1}^I \pi_{il}q_i \le F_l,\forall l \in \mathcal{L}\right\}$$

All the prosumers' updates are equivalent to:
\bsq
\label{eq:prosumer-update}
\begin{align}
    p^{k+1}=~&\mbox{argmin}\{\theta_2(p)-\frac{(b^k)^Tp}{a}  +\frac{||q^{k+1}+p-D||^2}{2a} \} \label{eq:prosumer-update.1}\\
    b^{k+1}=~&b^k-q^{k+1}-p^{k+1}+D \label{eq:prosumer-update.2}
\end{align}
\esq
where $\theta_2(p):=\sum \limits_{i=1}^I (c_ip_i^2+d_ip_i)-\frac{I-2}{2a(I-1)}\sum \limits_{i=1}^I (D_i-p_i)^2$.

If A2 holds, both $\theta_1(q)$ and $\theta_2(p)$ are convex, and the following function has a unique saddle point $(q^*,p^*,b^*)$.
\begin{align}
\label{eq:Lfuntion}
    \theta_1(q)+\theta_2(p)-\frac{b^T(p+q-D)}{a}
\end{align}

By variational inequality, \eqref{eq:platform-update.1} is equivalent to:
\begin{align}
    q^{k+1} \in \mathcal{Q},~\mbox{and for all}~q \in \mathcal{Q},~ \theta_1(q)-\theta_1(q^{k+1})   \nonumber\\ 
    +(q\!-\!q^{k+1})^T \left\{\!-\!\frac{1}{a}b^k+\frac{1}{a}(q^{k+1}\!-\!q^k\!+\!b^{k-1}\!-\!b^k)\right\} \ge 0 
\end{align}
By \eqref{eq:prosumer-update.2}, substitute $b^k$ with $b^{k-1}-q^k-p^k+D$:
\begin{align}
\label{eq:condition1}
    q^{k+1} \in \mathcal{Q},~\mbox{and for all}~q \in \mathcal{Q},~ \theta_1(q)-\theta_1(q^{k+1})   \nonumber\\ 
    +(q\!-\!q^{k+1})^T \left\{\!-\!\frac{1}{a}b^k+\frac{1}{a}(q^{k+1}+p^k-D)\right\} \ge 0  
\end{align}
Similarly, prosumers' update \eqref{eq:prosumer-update.1} is equivalent to:
\begin{align}
\label{eq:condition2}
p^{k+1} \in {\mathbb{R}^I},~\mbox{and for all}~p \in \mathbb{R}^I,~ \theta_2(p)-\theta_2(p^{k+1}) \nonumber\\
+ (p-p^{k+1})^T   \left\{-\frac{1}{a}b^k + \frac{1}{a}(q^{k+1}+p^{k+1}-D)\right\} \ge 0
\end{align}
Eliminating $b^k$ in \eqref{eq:condition1} and \eqref{eq:condition2} by \eqref{eq:prosumer-update.2}, we get:
\bsq
\label{eq:condition3}
\begin{align}
        q^{k+1} \in \mathcal{Q}, ~\mbox{and for all}~q \in \mathcal{Q},~ \theta_1(q)-\theta_1(q^{k+1})\nonumber \\
        +(q-q^{k+1})^T  \left\{-\frac{1}{a}b^{k+1}+\frac{1}{a}(p^k-p^{k+1})\right\} \ge 0 \label{eq:condition3.1}\\
        p^{k+1} \in \mathbb{R}^I, ~\mbox{and for all}~p \in \mathbb{R}^I,~ \theta_2(p)-\theta_2(p^{k+1}) \nonumber \\ 
        + (p-p^{k+1})^T    \left\{-\frac{1}{a}b^{k+1} \right\} \ge 0 \label{eq:condition3.2}
\end{align}
\esq
With $t:=(q,p)$ \footnote{For brevity, we use $(q,p) = [q^T, p^T]^T$ to represent a column vector.} and $\theta(t):=\theta_1(q) + \theta_2(p)$, \eqref{eq:condition3} implies:
\begin{align} 
    \theta(t)-\theta(t^{k+1})+ 
    \left(                
  \begin{array}{c}   
    q-q^{k+1} \\ 
    p-p^{k+1} \\
    b-b^{k+1} \\
  \end{array}
\right)^T \cdot ~& \nonumber\\
    \left\{\left(                
  \begin{array}{c}   
    -b^{k+1}/a \\ 
    -b^{k+1}/a \\
    q^{k+1}\!+\!p^{k+1}\!-\!D \\
  \end{array}
\right)+
 \left( 
  \begin{array}{c}   
    \mathbf{I}/a \\
    \mathbf{I}/a \\
    0\\
  \end{array}
\right)(p^{k}-p^{k+1})
 \right.  ~& \nonumber\\
\left.+
 \left(                
  \begin{array}{cc}   
    0 & 0 \\
    {\mathbf{I}/a} & 0 \\
    0 & \mathbf{I}/a\\
  \end{array}
\right)
 \left(                
  \begin{array}{c}   
    p^{k+1}-p^{k} \\
    b^{k+1}-b^{k} \\
  \end{array}
\right)\right\}  ~& \ge 0 \nonumber
\end{align}
for all $w:=(q,p,b) \in \mathcal{Q}\times \mathbb{R}^{2I}$.
Define the mapping $F(w):=(-b/{a},~-b/{a},~q+p-D)$, which is indeed monotone. Then:
\begin{align}
    \forall w \in {\mathcal{Q}\times \mathbb{R}^{2I}}: ~ \theta(t)-\theta(t^{k+1}) \nonumber\\
    +(w-w^{k+1})^T\left\{F(w^{k+1})+\left( 
  \begin{array}{c}   
    \mathbf{I}/a \\
    \mathbf{I}/a \\
    0\\
  \end{array}
\right)(p^{k}-p^{k+1})\right\} \nonumber\\
\ge \left( 
  \begin{array}{c}   
    p-p^{k+1} \\
    b-b^{k+1} \\
  \end{array}
\right)^T
\left( 
  \begin{array}{cc}   
    \mathbf{I}/a & 0 \\
    0 &  \mathbf{I}/a \\
  \end{array}
\right)
\left( 
  \begin{array}{c}   
    p^k-p^{k+1} \\
    b^k-b^{k+1} \\
  \end{array}
\right) \label{eq:8}
\end{align}

Recall $w^*:=(q^*,p^*,b^*)$ is the unique saddle point of \eqref{eq:Lfuntion}. Then by monotonicity of $F$, we have
\begin{align}
   & \theta(t^{k+1})-\theta(t^*)+(w^{k+1}-w^*)^TF(w^{k+1}) \nonumber\\
    \ge ~&\theta(t^{k+1})-\theta(t^*)+(w^{k+1}-w^*)^TF(w^*)  { \geq 0} \nonumber
\end{align}
and therefore \eqref{eq:8} implies:
\begin{align}
\label{eq:10}
 &   \left( 
  \begin{array}{c}   
    p^{k+1}-p^* \\
    b^{k+1}-b^* \\
  \end{array}
\right)^T
\left( 
  \begin{array}{cc}   
    \mathbf{I}/a & 0 \\
    0 &  \mathbf{I}/a \\
  \end{array}
\right)
\left( 
  \begin{array}{c}   
    p^k-p^{k+1} \\
    b^k-b^{k+1} \\
  \end{array}
\right) \nonumber\\
\ge ~ & (w^{k+1}-w^*)^T \left( 
  \begin{array}{c}   
    \mathbf{I}/a \\
    \mathbf{I}/a \\
    0\\
  \end{array}
\right)(p^{k}-p^{k+1}) \nonumber \\
=~& \frac{1}{a}(b^k-b^{k+1})^T(p^k-p^{k+1}) 
\end{align}
where the last equality utilizes \eqref{eq:prosumer-update.2} together with the saddle-point condition $q^*+p^*=D$.

Note that \eqref{eq:condition3.2} holds for $(p^k,b^k)$ and $(p^{k+1},b^{k+1})$. Making $p=p^{k+1}$ for the case with $(p^k,b^k)$ and $p=p^{k}$ for the case with $(p^{k+1},b^{k+1})$, and adding the two inequalities, we have:
\begin{align}
    \frac{1}{a}(b^k-b^{k+1})^T(p^k-p^{k+1}) \ge 0 \label{eq:11}
\end{align}

Combining \eqref{eq:10} and \eqref{eq:11}, we have:
\begin{align}
   \frac{1}{a}  \left( 
  \begin{array}{c}   
    p^{k+1}-p^* \\
    b^{k+1}-b^* \\
  \end{array}
\right)^T
\left( 
  \begin{array}{c}   
    p^k-p^{k+1} \\
    b^k-b^{k+1} \\
  \end{array}
\right) \ge 0 \nonumber
\end{align}
which further implies:
\begin{align}
\label{eq:converge}
  \bigg\| 
  \begin{array}{c}   
    p^k-p^{*} \\
    b^k-b^{*} \\
  \end{array}
\bigg\|^2 =~& 
   \bigg\| 
  \begin{array}{c}   
    p^{k+1}-p^{*} \\
    b^{k+1}-b^{*} \\
  \end{array}
\bigg\|^2 + 
\bigg\| 
  \begin{array}{c}   
    p^k-p^{k+1} \\
    b^k-b^{k+1} \\
  \end{array}
\bigg\|^2 \nonumber\\
  ~ & + 2 \left( 
  \begin{array}{c}   
    p^{k+1}-p^* \\
    b^{k+1}-b^* \\
  \end{array}
\right)^T
\left( 
  \begin{array}{c}   
    p^k-p^{k+1} \\
    b^k-b^{k+1} \\
  \end{array}
\right)  \nonumber\\
\ge ~ &  \bigg\|
  \begin{array}{c}   
    p^{k+1}-p^{*} \\
    b^{k+1}-b^{*} \\
  \end{array}
\bigg\|^2 + 
\bigg\|
  \begin{array}{c}   
    p^k-p^{k+1} \\
    b^k-b^{k+1} \\
  \end{array}
\bigg\|^2
\end{align}
The sequence $\{(p^k, b^k)\}$ is F\'ejer monontone, with $\|(p^k-p^*)^T, (b^k-b^*)^T\|^2$ decreasing in each iteration $k$ by $\|(p^k-p^{k+1})^T, (b^k-b^{k+1})^T\|^2$. As a result, the sequence $\{\|(p^k-p^*)^T, (b^k-b^*)^T\|^2\}$ converges and sequences $\{p^k\}$ and $\{b^k\}$ are bounded. With \eqref{eq:converge}, the sequence $\{p^k\}$ ($\{b^k\}$) only has one cluster point. According to \eqref{eq:condition3.2} we can get $p^k \to p^*$ and $b^k \to b^*$. Then with \eqref{eq:platform-update.1} we know $q^k \to q^*$ and $\lambda^k \to \lambda^*$.

\end{document}